\documentclass[aps, a4paper,superscriptaddress, nofootinbib,  11pt]{revtex4}

\usepackage{amsmath,amssymb,amsfonts, bm, natbib}
\usepackage{dsfont}
\usepackage{epsfig}
\usepackage{graphicx}
\usepackage{slashed}
\usepackage{color}
\usepackage{subfigure} 
\usepackage{caption}
\usepackage{slashed} 
\usepackage{graphicx}
\usepackage{dcolumn}
\usepackage{bm}

\newcommand{\be}{\begin{equation}}
\newcommand{\ee}{\end{equation}}

\newcommand{\wt}{\widetilde}

\newcommand{\bea}{\begin{eqnarray}}
\newcommand{\eea}{\end{eqnarray}}

\newcommand{\bfig}{\begin{figure}}
\newcommand{\efig}{\end{figure}}
\newcommand{\wh}{\widehat}

\newcommand{\gev}{\, \text{GeV}}

\makeatletter
\newcommand*{\rom}[1]{\expandafter\@slowromancap\romannumeral #1@}
\makeatother

\begin{document}

~\vspace{1cm}

\title{\LARGE {\bf \sffamily \boldmath  Perturbative Expansions in QCD  Improved  by Conformal Mappings of the Borel Plane}
\vspace{0.5cm}}

\author{Irinel Caprini\vspace{0.4cm}}
\affiliation{\it National Institute of Physics and Nuclear Engineering,
 Bucharest-Magurele, Romania\vspace{0.2cm}}
\author{Jan Fischer}
\affiliation{\it Institute of Physics, Academy of Sciences of the Czech Republic, 
Prague, Czech Republic\vspace{0.2cm}}
\author{Gauhar Abbas}\affiliation{\it Theoretical Physics Division, Physical Research Laboratory, Navrangpura, Ahmedabad, India\vspace{0.2cm}}
\author{ B. Ananthanarayan}\affiliation{\it Centre for High Energy Physics,
Indian Institute of Science, Bangalore, India\vspace{1.5cm}}

\vspace*{-0.7cm}
\begingroup
\let\newpage\relax
\endgroup

\begin{abstract}
\centerline{\bf Abstract \vspace{0.3cm}}
Perturbation expansions appear to
be divergent series  in many physically interesting situations, including in quantum field theories like quantum electrodynamics (QED) and quantum chromodynamics  (QCD), where the perturbative coefficients exhibit a factorial growth at large orders.   While this  feature 
has little impact on physical predictions in QED,  it can have nontrivial consequences in applications of perturbative QCD at moderate energies. In particular, it affects the   theoretical error in the extraction of the strong coupling $\alpha_s$ from hadronic  $\tau$  decays, despite progress of perturbative calculations available at present to four loops.  We discuss a new  type of perturbative expansion for QCD correlators, which uses instead of the standard powers of the  coupling  a new set of expansion functions. These functions are defined by means of an
optimal conformal mapping of the Borel complex plane, which implements the known features of the high-order divergence in terms of the lowest Borel-plane singularities. 
The  properties of the expansion functions resemble those of the expanded correlators, by exhibiting in particular the singular behaviour of the correlators at $\alpha_s=0$.  We prove the good convergence properties of the new expansions on mathematical models that simulate  the physical polarization function for light quarks  and its derivative (the Adler function), in various prescriptions of renormalization-group summation. 

\end{abstract}

\maketitle

\section{ Introduction}

Most of the problems in physics are plagued by the lack of exact solutions. To find a suitable approximation, one has to neglect a number of effects, thereby easing the labour but, simultaneously, endangering the physical relevance. It is  a wide experience that equations in physics can, as a rule, be solved only approximately.

    Perturbation theory (PT) is based on the idea of expressing the solution $F(z)$ of a problem in a (maybe formal) series in powers of a perturbative parameter $z$: 
\be\label{eq:f}
F(z) \simeq \sum_{n} f_n z^n,
\ee
where  $f_n$  are the expansion coefficients, and  $z$  is considered to be a small quantity  from which, however, significnt physical effects may result. 

       Two questions are of interest related to (\ref{eq:f}):

(i)    what is the meaning of the sum on the right-hand side of (\ref{eq:f}), and  

     (ii)   what is the meaning of the „equality“ sign in (\ref{eq:f}). 

The situation is simple when the infinite series in (\ref{eq:f}) is convergent for a certain value of  $z$. Then, at that particular value of $z$,  $F(z)$  is equal to the unique sum of the infinite series:
\be\label{eq:f1}
F(z)=\sum_{n=0}^\infty f_n z^n.
\ee        			
In general, let us assume  that $z$ is a complex variable.  If $F(z)$ is holomorphic inside a circle of radius $\rho>0$  centered at the origin, (\ref{eq:f}) represents  $F(z)$  uniquely in the form of the Taylor expansion of  $F(z)$ for all  $|z| < \rho$. The expansion coefficients $f_n$  are obtained from the derivatives to all orders of  $F(z)$  at  $z = 0$. 
For  $|z| > \rho$,  the power series in (\ref{eq:f}) is divergent and the sum is not defined. Finally, if the convergence radius $\rho$  is zero, the sum in (\ref{eq:f}) is not defined either and the equality (\ref{eq:f1}) can say nothing about anything related to interactions. 

If, however, (\ref{eq:f}) is understood, instead of (\ref{eq:f1}), as an asymptotic relation between  $F(z)$  and the sum, and we write
\begin{equation}\label{eq:f2}
F(z) \,\, \sim  \,\, \sum_{n=0}^{\infty} f_{n} z^{n},\quad\quad\quad  
 z \rightarrow 0,
\end{equation} 
then a function $F(z)$ may exist even if the series in (\ref{eq:f2}) is divergent. We recall that (\ref{eq:f2}) means that there exists a region ${\cal S}$ containing the 
origin or at least having it as an accumulation point, such that the set of functions 
\begin{equation}
R_{N}(z) = F(z) - \sum_{n=0}^{N}f_{n}z^{n} 
\label{rema}
\end{equation}
satisfy the condition 
\begin{equation}
R_{N}(z) = o(z^{N}) 
\label{ordo}
\end{equation}
for all $N=0,1,2,...$, $z \rightarrow 0$ and $z \in {\cal S}$ \cite{Hardy, Jeff}.

We emphasize that an asymptotic series is defined by a different limiting 
procedure than the Taylor series: taking $N$ fixed, one observes how 
$R_{N}(z)$ behaves for $z \to 0$, $z \in {\cal S}$,  the procedure being 
repeated for all $N \geq 0$ integers. In a Taylor series, $z$ is 
fixed and one observes how the sums $\sum_{n=0}^{N}f_{n}z^{n}$ behave for $N 
\to \infty$. Convergence, a property of the expansion coefficients $f_{n}$, 
may be provable without knowing the function $F(z)$ to which the series converges. 
However, asymptoticity can be tested only if one knows {\em both} 
the coefficients $f_{n}$ {\em and} the function $F(z)$.  
In contrast to (\ref{eq:f1}), the relation (\ref{eq:f2}) does not determine the function $F(z)$ uniquely, even if all the  coefficients $f_n$  are explicitly known and the set of rays approaching the origin $z = 0$  is specified.   

          Perturbative methods are used in astronomy, in quantum mechanics and in elementary particle physics, where the parameter $z$ measures the strength of particle interaction, while $z =0$  corresponds to the state when interaction is absent.  The applicability of perturbation theory is entirely dependent on the convergence properties of the power series (\ref{eq:f}), which are determined by the behavior of the large-order terms, and by the analyticity properties of the expanded function $F(z)$ at the expansion point $z=0$.  This has far-reaching consequences in quantum field theory (QFT).

\section{Divergent Perturbative Series in QFT}
Quantum field theories rely on two fundamental pillars of physics, quantum mechanics  and the special theory of relativity.  Notable examples are quantum electrodynamics (QED), 
an abelian  gauge theory which describes the electromagnetic interactions of quarks and leptons,  and quantum chromodynamics (QCD),
 a gauge theory based on SU(3) color group, which describes the strong interactions between the colored quarks
and gluons.  It enjoys the property of asymptotic freedom, if the number of fermion
families does not exceed a certain limit.  On the other hand, QCD is required
to be a confining theory as no free quarks and gluons are observed in nature. The gauge field theories  have been shown by 't Hooft and Veltman to be renormalizable, even if the symmetry is spontaneously  broken, as is the case of the unified theory of electromagnetic and weak interactions (the Glashow-Salam-Weinberg model).

Except for some idealized models, field theories
in general cannot be solved exactly. Perturbation theory is the basic tool for calculations: the physical quantities of interest,
such as scattering amplitudes, are expressed as perturbation series of the form (\ref{eq:f}) in powers of a renormalized coupling constant $z$, with coefficients $f_n$ obtained from the calculation of successive terms visualised by Feynman diagrams.

 In the case of QED,  the parameter $z$ in the perturbative expansion (\ref{eq:f}) is the fine structure constant $\alpha=e^2/4\pi$, where $e$ is the magnitude of the  electron charge. The coefficients $f_n$  have been calculated in some cases up to high perturbative orders. Examples are  the magnetic moments of the electron \cite{Aoyama:2012wj} and the muon \cite{Aoyama:2012wk}, for which  QED perturbation theory  makes predictions with an amazing accuracy, never reached before in science. 

In the case of QCD, the modern theory of strong interactions, the perturbative  parameter $z$ is the scale-dependent renormalized strong coupling $\alpha_s=g^2/4\pi$, where $g$ is the parameter entering the QCD Lagrangian. Perturbation theory is the basic tool for describing the quark and gluon jet production in high-energy processes and the influence of strong interactions on electroweak processes through higher order quantum fluctuations. It is valid on a wide range of energy scales, from very high energies down to several GeV.\footnote{At lower energies, where the strong coupling is no longer a small parameter, QCD perturbation theory is not applicable. In this range, effective field theories like chiral perturbation theory (ChPT) and nonperturbative approaches as lattice QCD  are the main tools for the study of strong interactions.}

It is, however,  known that the perturbative series in both QED and QCD are divergent series. The result obtained in 
1952 by Freeman Dyson for QED \cite{Dyson} was a surprise  and set a challenge for a radical reformulation of perturbation theory. 
Dyson's argument has been repeatedly critically discussed, reformulated and extended to other field theories including QCD  (see \cite{Lautrup}-\cite{Beneke} and references therein).

 The fact that the perturbative series in QFT are divergent can be inferred from two kinds of arguments: on the one hand, one can prove that the expanded functions (usually the Green functions of the theory) are singular at the expansion point, $z=0$. For QED this argument was used by Dyson \cite{Dyson}.   In the case of QCD, the argument is based on renormalization group invariance and was put forward by 't Hooft \cite{tHooft}. On the other hand, the divergency is inferred from studies of higher order terms of the series, based on Feynman diagrams, which indicate a factorial growth of the expansion coefficients, $f_n\sim n!$, for several field theories including QED and QCD.

 To give the divergent
series a precise meaning, Dyson proposed to interpret it as asymptotic to 
the desired function, {\em i.e.} he assumed that (\ref{eq:f2}) holds.
By this, the philosophy of perturbation theory changed radically.  Perturbation
theory yields, at least in principle, the values of all the $f_{n}$ coefficients. 
This can tell us whether the series is convergent or not. What we want to 
know is under what conditions the function $F(z)$ can be determined from (\ref{eq:f}).
If the series were  convergent, the knowledge of all the coefficients $f_{n}$ would uniquely determine $F(z)$.   
On the other hand, there are infinitely many functions having the same {\it 
asymptotic} expansion (\ref{eq:f2}).

It may seem surprising that the  field correlators have singularities in $z$ at the point $z=0$ which corresponds to the interaction vanishing. It is well known that interactions play a fundamental role in the formation of structures in Nature: the celestial bodies and all structures on the Earth exist due to the interaction of elementary particles. It is hardly imaginable what the Universe would be like without interaction: no forces, no structures, nothing but free particles, chaotic agglomerations, random multiplicities.
 
     The enormous difference between the world with interaction,  $z\ne 0$, and that without interaction,  $z=0$,  poses the question whether there is a physical relation between the two worlds. The great difference suggests that it would be unreasonable to try to explain the behavior of the interacting particles on the basis of the non-interacting ones. {\em i.e.}, to base the explanation of something existing (interaction) on something not existing (no interaction). The effect of interaction is  hidden in the singularity of $F(z)$ at $z=0$ or, if one insists in using power expansions,  in the derivatives  of $F(z)$ of all orders at the origin, which however do not exist. The singularity does not show up in the truncated low-order perturbative expansion which, being a polynomial in the parameter $z$, is holomorphic for any  $z$,  except for the point $z\to \infty$.  Therefore, going beyond finite orders is essential for capturing the essential properties of the theory. 

In our presentation we shall deal with these questions with specific reference  to QCD. We must emphasize that considerable progress has been achieved in perturbative QCD in the last decades: calculations to next-to-next-to-leading-order (NNLO) and next-to-next-leading-logarithm (NNLL) approximation are available for many high-energy processes and, as shown below,  for several observables the calculations have been pushed to even higher orders. However, the expansions are also plagued with some difficulties: the truncated, fixed order perturbative expansions are afflicted with  the problem of renormalization scheme and scale dependence and 
violate explicitly, due to the Landau singularities, the rigorous momentum-plane analyticity imposed by general principles of causality and unitarity on the correlations functions of the confined theory. Moreover, the perturbative expansions are not valid  in the kinematical regions where hadron interactions are measured,  an analytic continuation from the  euclidean to minkowskian regions being necessary for comparison with experiment. Finally, the ambiguities related to the fact that the expansions are divergent series have a much larger effect than for QED, due to the fact that at moderate energies, of a few GeV, the coupling is relatively large. As we shall discuss below, these difficulties are to a certain extent interconnected.

\begin{figure}[htb]\vspace{0.5cm}
\begin{center} \includegraphics[width =8cm]{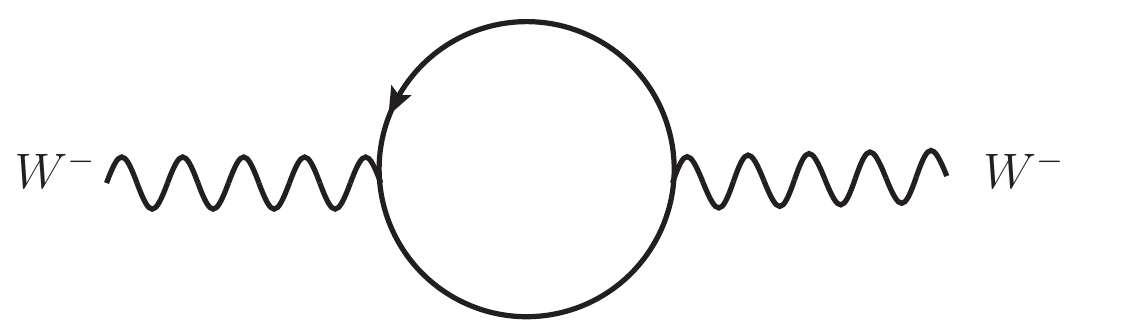}
\caption{The tensor (\ref{Pi}) to leading order. The solid lines denote light quarks. \label{fig:Pi}}\end{center}
\end{figure}

We consider for illustration the Adler function  in
massless QCD, defined as
\begin{equation}\label{D}
D(s) = - s \,\frac{d \Pi (s)}{ds}\,,
\end{equation}
where $\Pi (s)$ is the amplitude of the current--current correlation tensor
{\footnotesize\be\label{Pi}
 \Pi^{\mu\nu}(p)= i \int d^4 x\,e^{-ip x} 
\langle 0 |{\rm T} (j^{\mu}(x) j^{\nu}(0))| 0 \rangle =(g^{\mu \nu}p^2 - p^{\mu}p^{\nu}) \Pi (s),\quad\quad s=p^2,
\ee}
corresponding to a vector or an axial-vector current  $j^{\mu}$ of massless quarks (see Fig. \ref{fig:Pi}).

The function $D(s)$ is renormalization-group invariant and ultraviolet finite.  It can be calculated in perturbative QCD by inserting gluon and quark lines in the Feynman diagram (\ref{fig:Pi}).  Its formal perturbative expansion  reads
\begin{equation}
\label{Ds}
D_{\rm pert}(s) =1+ \sum\limits_{n\ge 1} \,\left(\frac{\alpha_s(\mu^2)}{\pi}\right)^n\,
\sum\limits_{k=1}^{n} k\, c_{n,k}\, (\ln (-s/\mu^2))^{k-1},
\end{equation}
where $\alpha_s(\mu^2)$ is the renormalized strong coupling at an arbitrary scale $\mu^2$. In particular, choosing $\mu^2=-s$,  (\ref{Ds}) takes the simple form
\be\label{DsRG}
D_{\rm pert}(s) = \sum\limits_{n\ge 0} \, c_{n,1} \left(\frac{\alpha_s(-s)}{\pi}\right)^n, 
\ee
where we denoted for convenience $c_{0,1}=1$. The series (\ref{DsRG}) is known as ``renormalization-group improved expansion'', because it avoids the appearance of large logarithms of the form $\ln(-s/\mu^2)$ in the coefficients,  the entire energy dependence being included in the  coupling. 

 The dependence of the coupling on the scale is governed by the 
renormalization-group equation
 \begin{equation}\label{eq:rge}
 \mu^2\frac {d\alpha_s (\mu^2)}
{d\mu^2}=\beta(\alpha_s)\equiv- \alpha_s (\mu^2)\sum_{n\ge 0}
\beta_n (\alpha_s (\mu^2))^{n+1}\,, \end{equation}
where the coefficients $\beta_n$ are calculated perturbatively and depend on the renormalization scheme for $n\ge 2$. At one-loop, this equation has the well-known solution\footnote{It is easy to see that the one loop coupling has a pole at a finite spacelike value $s=-\Lambda^2$. This pole, present also in the exact solution of (\ref{eq:rge}), produces an unphysical singularity in the truncated renormalization-group improved expansion (\ref{DsRG}), known as ``Landau pole''.} 
\be\label{a1loop}
\tilde a_s(-s)=\frac{a_s(\mu^2)}{1+\beta_0 a_s(\mu^2) \ln(-s/\mu^2)}\,,
\ee
where $ a_s\equiv \alpha_s/\pi$.  This equation exhibits asymptotic freedom,  $\tilde a_s(-s)\to 0$ for $s\to -\infty$. At two-loop, the solution of (\ref{eq:rge}) is expressed in terms of a Lambert function, while at higher orders the renormalization group equation can be integrated only numerically. 

The perturbative coefficients  $c_{n,k}$ for $n\ge 1$ in (\ref{Ds}) include the efect of higher-order quantum fluctuations. Actually, only the leading coefficients  $c_{n,1}$ require the evaluation of Feynman diagrams, the remaining ones, $c_{n,k}$ with $k>1$ are obtained in terms of  $c_{m,1}$ with $m< n$  and the coefficients $\beta_n$ of the $\beta$ function by imposing renormalization-group invariance to each order. 

The state-of-the-art is that $\beta$ function (\ref{eq:rge}) was calculated to five loops in  $\overline{{\rm MS}}$  scheme (see  \cite{BCK16} and references therein). For $n_f=3$ flavours the expansion coefficients are
\be\label{betaj}
 \beta_0=9/4,\,\, \beta_1=4,\,\, \beta_2= 10.0599,\,\,\beta_3=47.228,\,\,\beta_4=134.08.
\ee
The Adler function itself was calculated to four loops, which makes it one of the most precisely known Green functions in QCD. The leading coefficients $c_{n,1}$ in the $\overline{{\rm MS}}$-renormalization scheme with $n_f=3$ have the values (see \cite{BCK08} and references therein):
\be\label{cn1}
c_{1,1}=1,\,\, c_{2,1}=1.640,\,\, c_{3,1}=6.371,\,\, c_{4,1}=49.076.
\ee
On the other hand, for large $n$  the coefficients $c_{n,1}$ exhibit a generic factorial growth of the form \cite{Beneke}
\begin{equation}\label{D_n}
c_{n,1}\approx K\,b^n n !\, n^{c},\quad \quad n\to\infty,
\end{equation}
where $K$, $b$ and $c$ are  constants.
 Therefore, the  radius of convergence of the expansion (\ref{Ds}) is zero. This is related to the fact that the Adler function, viewed as a function of the strong coupling, is singular at the origin of the complex $\alpha_s$ plane. Furthermore, as shown by 't Hooft \cite{tHooft}, $D$  is analytic only in a horn-shaped region in the half-plane $\text{Re}\, \alpha_s>0$, of  zero opening angle near $\alpha_s=0$. 

\section{\bf Borel Summation}
 Several mathematical techniques for the summation of divergent power series are known, which under certain conditions recover the expanded function from its expansion coefficients \cite{Hardy}. For instance, the Borel summation  has received much interest in recent years and has been adopted for the summation of the perturbation series in QCD, although the mathematical conditions required for its use are not satisfied in this case. To illustrate the method, we  start from the expansion (\ref{DsRG})  of the Adler function  and define its Borel transform $B(u)$ by the series:
\be
\label{Borel}
B(u)= \sum_{n=0}^\infty b_n u^n,\quad\quad b_n= \frac{c_{n,1}}{\beta_0^n \,n!}\,.
\ee
One can check that the function $D(s)$ can be written formally in terms of $B(u)$ by means of the Laplace-Borel representation 
\begin{equation}\label{Laplace}
   D(s) =  \frac{1}{\beta_0 a_s(-s)}\,
    \int\limits_0^\infty\!{\rm d}u\, B(u)\,\exp
    \left(-\frac{u}{ \beta_0 a_s(-s)}\right) \,.
\end{equation}

Due to the $n!$ in the denominator of $b_n$, the series (\ref{Borel}) is expected to be convergent in a disk, $|u|<u_0$ of positive radius, $u_0>0$. If the function $B(u)$ could be analytically continued in the $u$ complex plane outside this disk up to the real axis, and the integral (\ref{Laplace}) were convergent for a certain $a_s>0$, then the original series would be Borel summable and (\ref{Laplace}) would define uniquely a function analytic in a region of the half-plane $\text{Re}\, a_s>0$.

 Criteria for Borel summability  are formulated  as constraints on the properties of the expanded function $D$  in the complex $a_s$ plane (for a review see \cite{Jan, Jan1}). Watson theorem \cite{Watson} requires the analyticity of $D$  in a region of the $a_s$ plane defined by $|a_s|<R$ and  $|\arg(a_s)|<\pi/2+\epsilon $, for certain positive numbers $R$ and $\epsilon$. A generalization is Nevanlinna criterion \cite{Nevanlinna}, which replaces the  sector $|\arg(a_s)|<\pi/2+\epsilon $ by the region $\text{Re}\, (1/a_s)<1/\eta$,  for some $\eta>0$.

\begin{figure}[htb]
\begin{center} \includegraphics[width = 9cm]{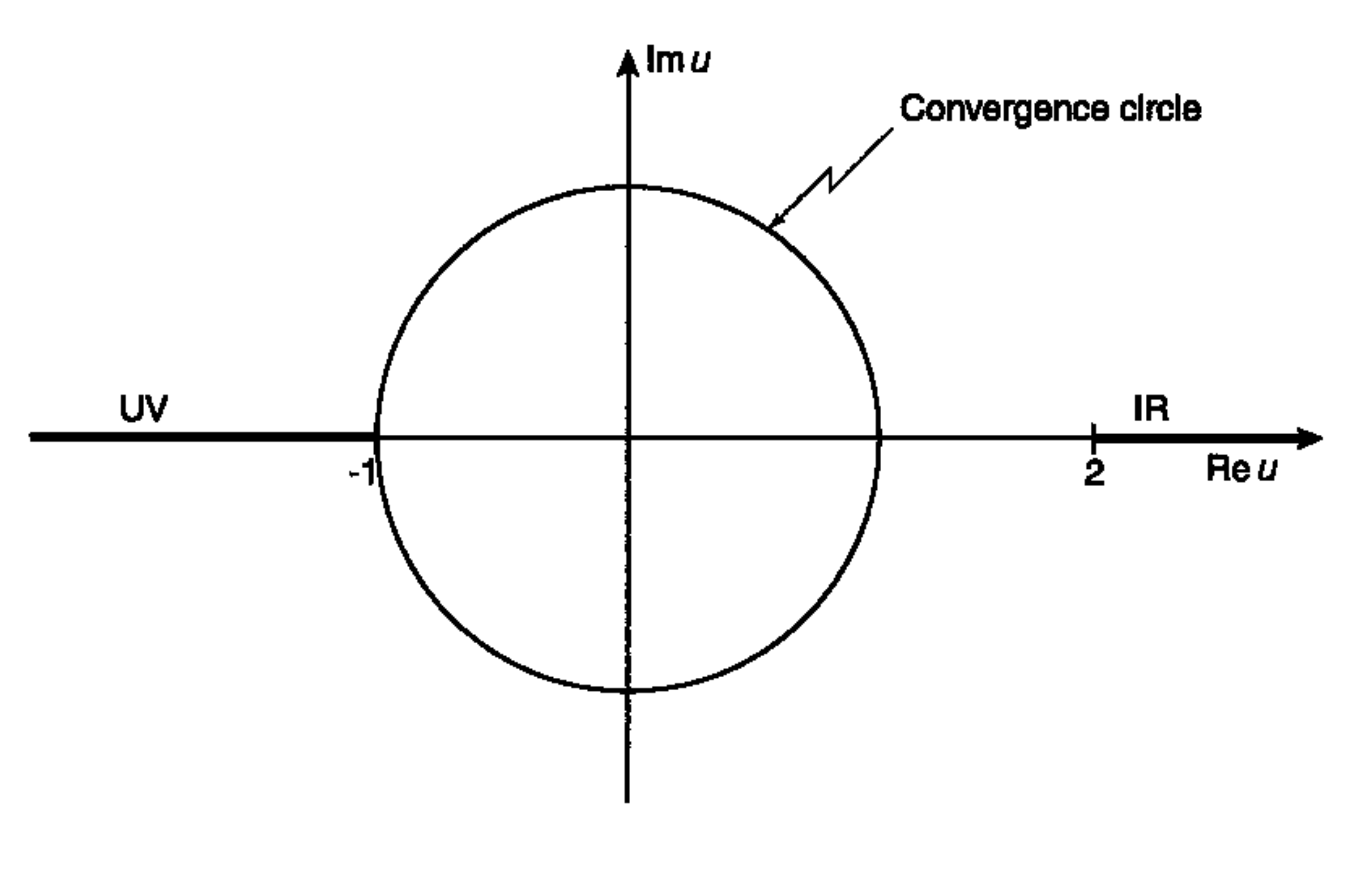} 
\caption{The Borel plane for the Adler function. The series (\ref{Borel}) converges inside the circle passing through the first UV renormalon at $u=-1$. \label{fig:uplane}}\end{center}
\end{figure}

These conditions are, however, not fulfilled in QCD, since the horn-shaped analyticity region found by 't Hooft  violates the Watson and Nevanlinna criteria. Alternatively, Borel non-summability results from the singularities of the Borel transform $B(u)$ in the $u$ plane: detailed studies \cite{Mueller1985, Beneke} have showed that $B(u)$ has singularities on the semiaxis $u\ge 2$, denoted as  infrared (IR) renormalons, and for $u\le -1$, denoted as ultraviolet (UV) renormalons. The names indicate the regions of the Feynman integrals, which are responsible for the appearance of these singularities.  Other singularities, at larger values on the positive real axis, are due to specific field configurations known as instantons.  Apart from  the two cuts along the lines $u\geq 2$ and $u\leq -1$, it is assumed that no other singularities are  present in the complex $u$ plane \cite{Parisi, Mueller1985}. The cut $u$ plane is shown in Fig. \ref{fig:uplane}.

We emphasize that the Borel transform encodes the large-order increase of the coefficients in its singularities  in the complex $u$ plane.  A first consequence is that, due to the singularities of $B(u)$ for $u\ge 2$, the Laplace-Borel integral (\ref{Laplace}) is not defined and is ambiguous.  In order to recover the original function $D(s)$, a prescription of regulating the integral is necessary.  The principal value (PV) prescription, the most natural choice for mathematicians, has been adopted also for summation of perturbative QCD \cite{Mueller1992, Beneke}. It is defined as
\begin{equation}\label{PV}
  \text{PV}  \int\limits_0^\infty\!du\, B(u)\,e^{-u/a}
    =\frac{1}{2}\left[
    \int_{\cal{C}_-}\!d u\, B(u)\,e^{-u/a}  +
    \int_{\cal{C}_+}  \! du\, B(u)\,e^{-u/a} \right], 
\end{equation}
 where ${\cal C_+}$ (${\cal C_-}$) are lines parallel  to the
real positive axis, slightly above (below) it and we denoted $a\equiv\beta_0 a_s$.
 As discussed in \cite{CaNe}, the PV prescription  is preferred from the point of view of the momentum-plane analyticity properties that must be satisfied by the QCD correlation functions.

If one adopts a certain prescription (e.g., the principal value prescription), it is possible to exploit the available knowledge of the large-order behavior of the coefficients for defining a new expansion, in which the divergent pattern is considerably tamed. Such an approach uses techniques of convergence acceleration based on ``conformal mappings'' and ``singularity softening'',  which will be explained in the next sections.

 Before ending this section, we want to mention an important consequence of the intrinsic ambiguity of perturbative QCD due to the IR singularities of the Borel transform. We note that an IR renormalon at  $u=k$, where  $k\ge 2$ is a positive integer, generates an ambiguity of the form $\exp(-k/\beta_0 a_s)$ in the integral (\ref{Laplace}). By using the one-loop expression (\ref{a1loop}) of the running coupling, and denoting $Q^2=-s$,  this is equivalent to an ambiguity of the form $1/Q^{2 k}$. Thus, the divergence and Borel non-summability of the QCD perturbative series  implies the existence of additional terms in the representation of the QCD correlators, consisting actually of  a whole series of power corrections \cite{Mueller1992}. These terms are alternatively inferred from the philosophy of operator product expansion (OPE) and reflect the properties of the QCD vacuum \cite{SVZ1,SVZ2}. The conclusion is that, besides the pure perturbative part $\Pi_{\rm pert}(s)$  obtained from the Adler function $D_{\rm pert}(s)$  using (\ref{D}), the correlator $\Pi(s)$ contains a whole series of power corrections  \cite{SVZ1,SVZ2}
\be\label{eq:OPE}
\Pi(s)\sim \Pi_{\rm pert}(s) + \Pi_{\rm PC}(s),
\ee
where
\be\label{eq:PCseries}
\Pi_{\rm PC}(s)\sim \sum_{n\ge 1} \frac{d_n}{Q^{2n}},\quad\quad \quad  Q^2=-s,
\ee
the coefficients $d_n$ being expressed in terms of factors calculated perturbatively and vacuum expectation values of higher-dimensional ($d>0$) quark and gluon operators  (the so-called ``vacuum condensates''). These quantities should be calculated using the same prescription as that adopted for the pure perturbative part $\Pi_{\rm pert}(s)$.
\section{Method of Conformal Mapping}
A conformal mapping or transformation in simple terms transforms two oriented intersecting curves from one complex plane to another complex plane, such that it preserves the angle between them in magnitude and in orientation.  This means that  the angle between two curves in the original plane will be identical to that of the angle between corresponding curves in  the second plane, although the transformed curves in the  latter plane may not be similar to the original curves in the  first plane.  A holomorphic function $F(z)$  is conformal at every point $z_0$ where $F^\prime  (z_0) \neq 0$.

The conformal mapping method was  introduced  in particle physics in Refs. \cite{CiFi, Frazer, CCF} for improving the convergence of the power series used  for the representaton of scattering amplitudes.   By this method, a series in powers of a certain variable, convergent in a disk of positive radius around the origin, is replaced by a series in powers of another variable, which actually performs the conformal mapping of the original complex plane (or a part of it) onto a disk of radius equal to unity in the transformed plane. The new series converges in a larger region,  well beyond the disk of convergence of the original expansion, and also has an increased  asymptotic convergence rate at points lying inside this disk. An important result proved in Refs. \cite{CiFi, CCF} is that the asymptotic convergence rate is maximal if the new variable maps the entire holomorphy domain of the expanded function onto the unit disk. This particular conformal mapping is called ``optimal''. 

For QCD, it turns out that the method  is not applicable to the 
formal perturbative series of $D$ in powers of  $\alpha_{s}$, because $D$  is singular 
at the point of expansion\footnote{In the so-called "order-dependent" conformal mappings, which were defined also in the coupling plane \cite{ZJ, ZJ1}, the singularity is  shifted away from the origin by a certain amount at each finite-order, and tends to the origin only when an infinite number of terms are considered.}.  However, the method can be applied, rather than to   $D(s)$,  to its Borel transform $B(u)$, which is holomorphic in a region containing the origin  $u = 0$  of the Borel complex plane and  can be expanded in powers of  the Borel variable as in (\ref{Borel}).

\begin{figure}[htb]
\begin{center}
 \includegraphics[width = 5.1cm]{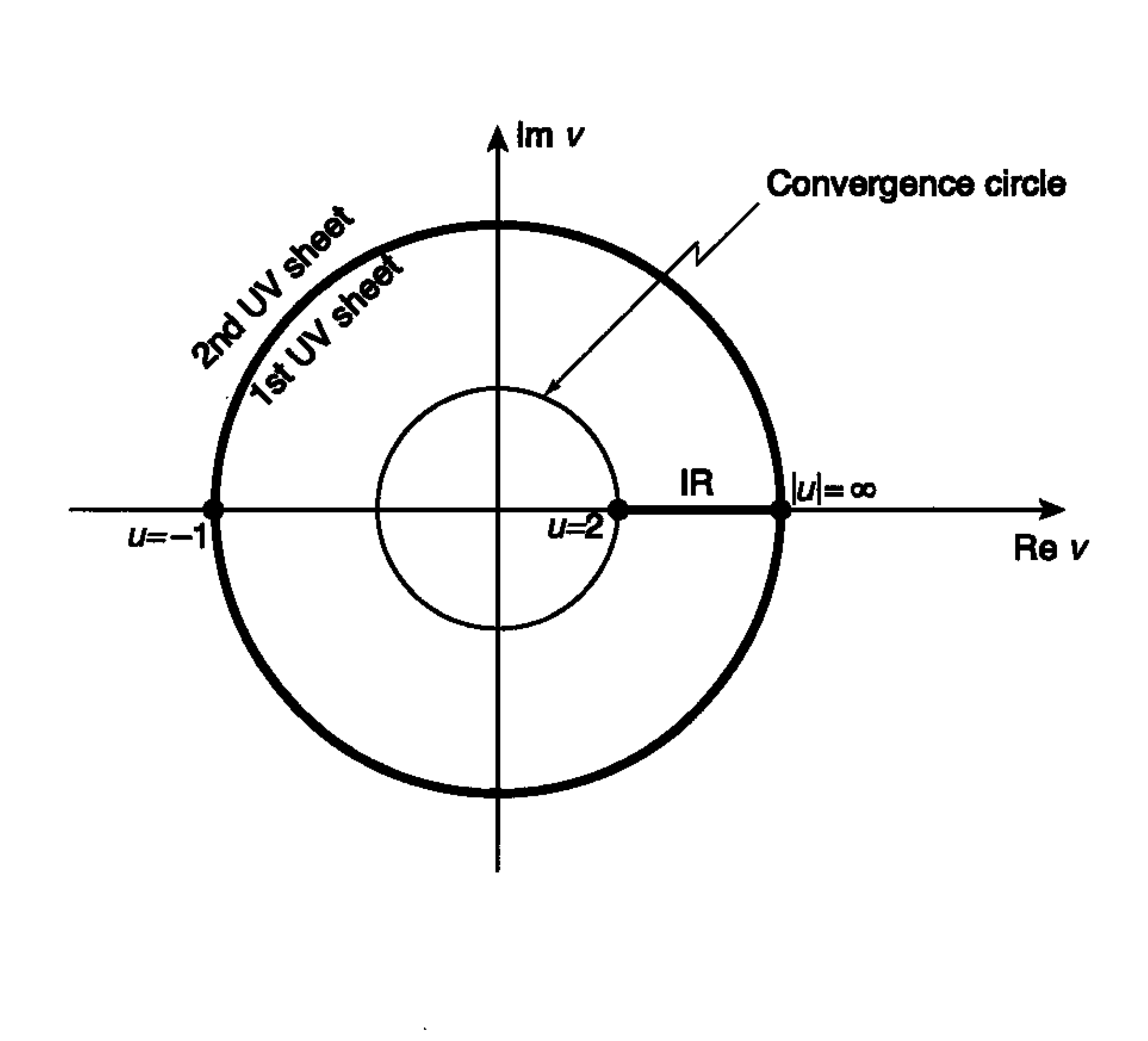} \includegraphics[width = 6cm]{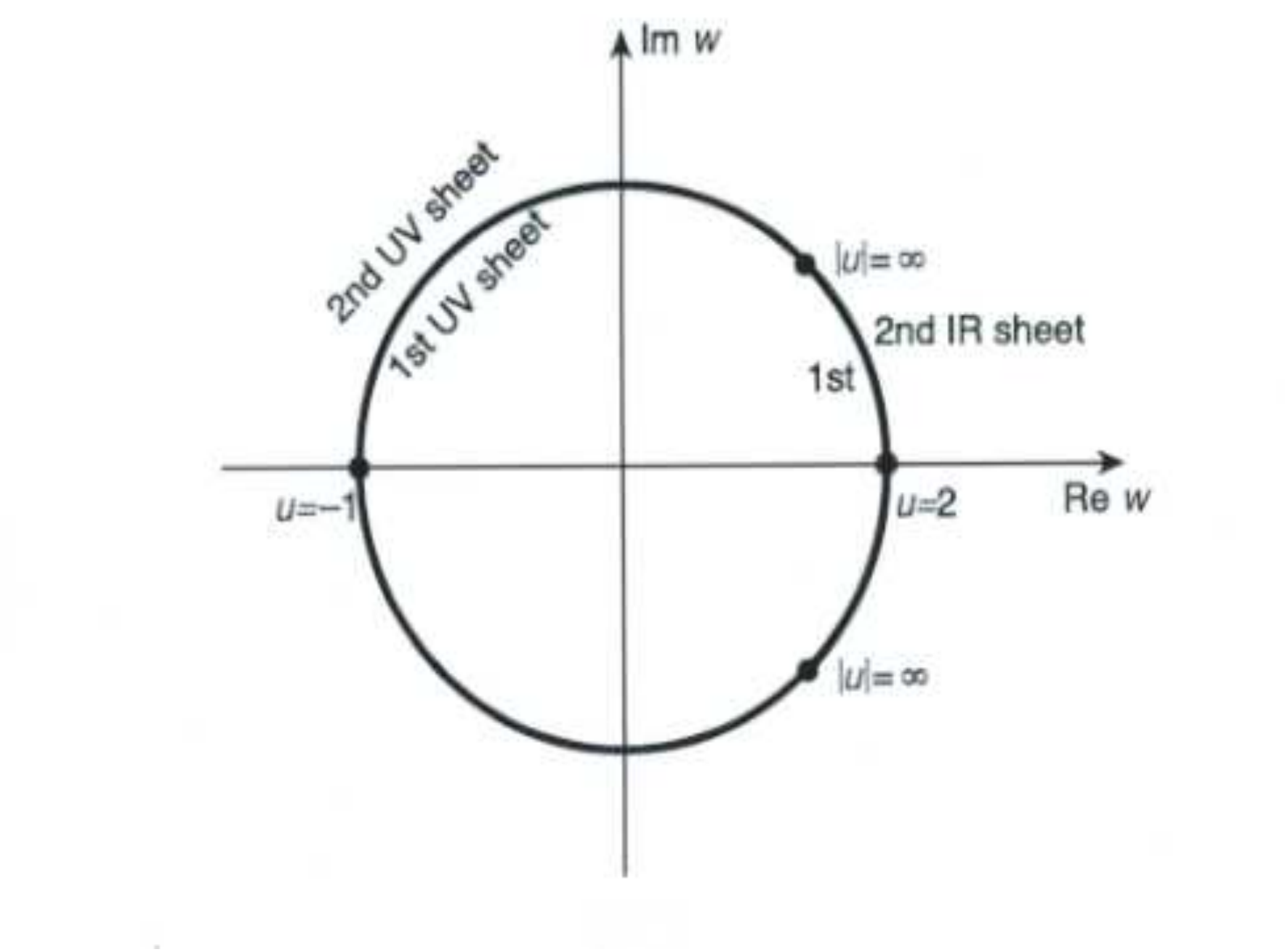}
\end{center}
\caption{Left: the $v$ complex plane. The UV cut is mapped upon the unit
circle, while the IR cut is situated inside it. The convergence domain of the
series (\ref{Bv}) is limited by  the image of the point
$u=2$, the lowest branch point of the IR cut. Right: the $w$ complex plane. Both the UV and IR cuts are mapped
on the boundary of the unit circle.  The
convergence domain is the whole Borel plane cut for $u\ge 2$ and $u\leq -1$.\label{fig:vw}}
\end{figure}

The conformal mapping of the Borel plane was suggested in
\cite{Mueller1992} as a technique to reduce or eliminate the ambiguities (power 
corrections) due to the large momenta in the Feynman integrals. As shown in Fig. \ref{fig:uplane}, the first UV renormalon at $u=-1$ limits the convergence of the series (\ref{Borel}) to the disk $|u|<1$, and generates therefore an ambiguity of the form $\exp(-1/\beta_0 a_s)$ in the Laplace-Borel integral (\ref{Laplace}). By using the argument presented above,  this is equivalent to an ambiguity of the form $1/Q^2$. However, this power correction is not a genuine ambiguity for QCD, because it is produced by large momenta in the  Feynman integrals, which are harmless. The spurious ambiguity can be eliminated by expanding $B(u)$ in a power series which converges also for $u>1$. This is achieved by the conformal mapping 
\begin{equation}v\equiv \tilde v(u) = \frac{\sqrt{1+u}-1}{\sqrt{1+u}+1} \,, \label{v}
\end{equation}
proposed by Mueller \cite{Mueller1992}  and used also in Refs. \cite{Alta, SoSu}. As one can see from  Fig. \ref{fig:vw} left, the function $\tilde v(u)$  maps the $u$ plane cut along the line $u\leq -1$ onto the unit disk $|v|<1$ in the $v$ plane. In the $v$ plane, the origin $u=0$ of the $u$ plane becomes the origin $v=0$,  the upper and lower edges of the cut $u\leq -1$ become the circle $|v|=1$, and the IR renormalon cut along $u\ge 2$ becomes a real segment inside the circle.
The corresponding expansion 
\begin{equation}
B(u)=\sum\limits_{n=0}^\infty d_n v^n\,,
\label{Bv}
\end{equation}
will converge in the disk limited by the image $\tilde v(2)$ of the first IR renormalon (see Fig. \ref{fig:vw} left).  This domain is larger than the original disk in Fig. \ref{fig:uplane}, but does not cover the entire $u$ plane. 
The reason is the fact that the conformal mapping  (\ref{v}) exploits only in part  the known
singularity structure in the Borel plane and is not optimal in the sense explained above.

 An optimal mapping, which performs 
the  analytic continuation in the entire doubly-cut Borel plane, was proposed for the first time
in \cite{CaFi1998} and was further investigated in \cite{CaFi2000, CaFi2001, CaFi2009, CaFi_RJP, CaFi2011, CaFi_Manchester} (similar methods were applied also in \cite{Jent1, CvLe}).  By means of this technique, it is possible to define a  non-power perturbative expansion in QCD in terms
of a new set of functions that fully exploit the location of the singularities
in the Borel plane.

As shown in \cite{CaFi1998}, the optimal conformal mapping of the plane $u$ for the Adler function is: 
\begin{equation}\label{eq:w}
w\equiv \tilde w(u)=\frac{\sqrt{1+u}-\sqrt{1-u/2}}{ \sqrt{1+u}+\sqrt{1-u/2}}\,.
\end{equation}
One can check that (\ref{eq:w})  maps the complex  $u$ plane cut along the real axis for $u\ge 2$ and $u\le -1$ onto the interior
of the circle $\vert w\vert\, <\, 1$ in the complex $w$-plane such that  the origin $u=0$ of the $u$ plane
corresponds to the origin $w=0$ of the $w$ plane, and the upper (lower) edges of the cuts are mapped onto the upper
(lower) semicircles in the  $w$ plane (see  Fig. \ref{fig:vw} right). The
 inverse of the mapping (\ref{eq:w}) is
\begin{equation}\label{uw}
u\equiv \tilde u(w)=\frac{8 w}{ 3 w^2-2 w +3}= \frac{8 w}{ 3 (w-\zeta) (w-\zeta^*)}\,,
 \end{equation}
where $\zeta= (\sqrt{2}+i)/(\sqrt{2}-i)$ and its complex conjugate  $\zeta^*$
are the images of $u=\infty$ on the unit circle in the $w$ plane.

By the  mapping (\ref{eq:w}), all  the singularities of the Borel transform, the  UV and IR  renormalons, have been pushed on the boundary of the unit disk in the $w$  plane, all at equal distance from the origin. Consider now the expansion of $B(u)$ in powers of the variable $w$: 
\begin{equation}\label{Bw}
B(u)=\sum_{n=0}^\infty c_n\,w^n\,,
\end{equation}
 where the coefficients $c_{n}$ can be obtained from  the coefficients $b_{k}$,
$k\leq n$, using Eqs. (\ref{Borel}) and  (\ref{eq:w}).  
By expanding $B(u)$ according to (\ref{Bw}) one makes full use of its
holomorphy domain, because the known part of it
({\em i.e.} the first Riemann sheet) is
mapped onto the convergence  disk. 

As we mentioned above,  an important result proved in \cite{CiFi}  is that the  expansion in powers of the optimal conformal mapping has  the
fastest asymptotic (large-order) convergence rate, compared to any other
expansion in powers of a variable that maps only a smaller part of the
holomorphy domain onto the unit disk. We recall		
that the large-order convergence rate of a power series is equal to that of 
the geometrical series with the quotient $r/R$, 
$r$ being the distance of the point from the origin and $R$ the convergence 
radius. 
The  proof given in \cite{CiFi} consists in comparing the magnitudes of the 
ratio  $r/R$ 
for a certain point in different complex planes, corresponding to different
conformal mappings. When the whole analyticity domain ${\cal D}$  of the
function is mapped on a disk, the value of $r/R$ is minimal \cite{CiFi}.  For a detailed proof, see
Ref.~\cite{CaFi2011}. 

 The expansion (\ref{Bw}) of the Borel transform suggests  an expansion for  the Adler
function of the form \cite{CaFi1998, CaFi2000, CaFi2001}
\begin{equation}
D(s)= \sum_{n=0}^{\infty} c_{n} W_{n}(a), 
\label{cW}
\end{equation}
where the functions $W_n(a)$ are defined as Borel-Laplace transforms of the integer powers of $\tilde w(u)$:
\begin{equation}\label{Wn}
W_{n}(a)=\frac{1}{ a}\int\limits_0^\infty\, e^{-u/a}\, (\tilde w(u))^n \,du, \quad\quad a\equiv\beta_0 a_s(-s).
\end{equation}
At each finite truncation order $N$, the expansion (\ref{cW}) is obtained
 by inserting the series (\ref{Bw})  into the Laplace integral (\ref{Laplace})
and exchanging the order of summation and integration. This procedure is
trivially allowed at any finite integer $N \geq 0$. For $N\to\infty$, however, 
the new expansion  (\ref{cW}) represents  a nontrivial step out of 
perturbation theory, replacing  the perturbative powers
$a^n$ by the functions $W_{n}(a)$. 

This procedure is an obvious generalization of the conformal mapping method 
proposed in \cite{Zinn}  for Borel-summable functions.  Formally, the expansion (\ref{cW}) is obtained
from the standard  perturbative expansion (\ref{DsRG}) by replacing the
coefficients $b_n$, appearing in the Taylor series (\ref{Borel}), by the
coefficients $c_n$ of the improved expansion (\ref{Bw}), and the
perturbative functions  $n! a_s^n$ (which multiply  the coefficients
$b_n$) by the new functions $W_{n}(a)$ defined  by the integral (\ref{Wn}).

\section{Properties of the New Expansion Functions}

We note first that the integral (\ref{Wn}) is not well-defined, since the variable $w=\tilde w(u)$ has a branch point singularity at the point $u=2$, which is situated along the integration range. This is a manifestation of the
intrinsic ambiguity of the perturbation theory produced by the infrared
renormalons.  According to the discussion above, a
prescription is required for defining the integral, which we take to be the same PV prescription (\ref{PV}) adopted for the correlator $D$
itself.  So, we shall define
\be\label{pv} 
W_{n}(a)= \frac{1}{ a}{\rm PV} \int\limits_{0}^\infty  e^{-u/a}\,  (\tilde w(u))^n\,du.
\ee 
In what follows we
shall briefly discuss the properties of the  expansion functions $W_n(a)$, showing that in many respects
they resemble the expanded function  $D(s)$ itself.

A first question is what are the analyticity properties of the expansion functions in the complex $a$ plane (we recall that $a$ is related to the strong coupling by $a=\beta_0 \alpha_s(-s)/\pi$). The problem of the analytic properties of the QCD correlators in the coupling
constant plane is very complicated. 't Hooft \cite{tHooft} and
Khuri \cite{Khuri1} showed that renormalization group invariance and the
multiparticle branch points on the timelike axis of the $s$ plane 
 imply a complicated
accumulation of singularities near the point $a=0$. Since the proof uses a
nonperturbative argument (multiparticle states generated by confinement in
massless QCD),  it is difficult to see this feature in standard truncated  perturbation theory: indeed,   the standard expansions in powers of $a$, truncated at a finite order, are holomorphic at $a=0$ and cannot capture this property of the full correlator.

For the new expansion functions  $W_{n}(a)$, from their definition (\ref{pv}) one can expect a more complex structure in the $a$ plane, even after the regularization of the integral by the PV prescription. The detailed analysis performed in \cite{CaFi2001} shows that the functions $W_{n}(a)$ are 
analytic functions of real type, {\em i.e.} they satisfy the Schwarz reflection property  $W_{n}(a^*)=  (W_{n}(a))^*$, in the whole complex $a$ plane, except for a cut along the
real negative axis and an essential singularity at $a=0$. Thus, even a truncated expansion (\ref{cW}) will exhibit a feature of the full correlator, namely its singularity at the origin $a=0$, although the exact nature of the singularity can not be captured.

It is of interest to investigate also the perturbative expansion of the
functions $W_n$ in powers of $a$. Since
$W_{n}(a)$ have singularities at $a=0$,  their Taylor
expansions around the origin will be  divergent series. We take first $a$
real and positive. The asymptotic  expansion is obtained by applying  Watson's
lemma \cite{Watson1918} (see also \cite{Jeff} and \cite{CaFiWatson}).

Specifically, we consider 
the Taylor expansion \begin{equation}\label{wnser}
(\tilde w(u))^n=\sum\limits_{k=n}^\infty \xi_k^{(n)} u^k\,,
\end{equation}
which is convergent for $|u|<1$. The sum begins with $k=n$ since, as follows
from (\ref{eq:w}),  the derivatives $(\tilde w^{n})^{(k)}(0)$ vanish for $k<n$ (in
particular $\xi^{(n)}_n=(3/8)^n$).  Then one can prove the relation \cite{CaFi2001}
 \be\label{Was}
W_{n}(a) = \sum\limits_{k=n}^N \xi_k^{(n)} k! a^k + {\tilde
M}_n \,(N+1)! \,a^{N+1} + O\left({\rm e}^{-\frac{X}{ a}}\right)\,,\nonumber
\ee 
where $N$ is a positive integer, ${\tilde M}_n$ is independent of $N$ and $X$ is an arbitrary positive parameter less than 1. From the definition (\ref{eq:f2}), it follows that  $W_{n}(a)$ admit the asymptotic
series   \begin{equation}\label{Was0}
W_{n}(a) \sim \sum\limits_{k=n}^\infty \xi_k^{(n)} k! a^k \,,\quad a\to 0_+\,\,.
\end{equation} 
The expansion (\ref{Was0}) is independent of the prescription
required in the definition of $W_n(a)$. We note that the first term of each $W_{n}(a)$ is
proportional to $n! a^n$ with a positive coefficient,  thereby  retaining a
fundamental property of perturbation theory. But the series (\ref{Was0}) 
are divergent: indeed, since the expansions (\ref{wnser}) have their convergence
radii equal to 1, then for any $R > 1$ there are infinitely many $k$ such that  
$|\xi^{(n)}_k|>R^{-k}$ \cite{Jeff}. Actually, the divergence of the series 
(\ref{Was0}) is not surprising, in view of the singularities of the functions  $W_{n}(a)$ at the origin of the $a$ plane. 

 For illustration we
give below the expansions of the first functions $W_n(a)$, derived in \cite{CaFi2001}:
{\small \begin{eqnarray} W_{1}(a)
&\sim& 0.375 a - 0.187 a^{2} + 0.457 a^{3}- 1.08 a^{4} + 4.32
a^5+\ldots, \nonumber \\ W_{2}(a) &\sim& 0.281 a^{2}- 0.422 a^{3} + 1.58 a^{4}
- 5.80 a^{5} + 29.78 a^6+\ldots, \nonumber \\  W_{3}(a)&\sim& 
0.316 a^{3}- 0.949 a^{4}+ 5.04 a^{5}- 25.95 a^6 +167.99
a^7+\ldots  \label{sim} \end{eqnarray} }
The higher powers of $a$  become quickly
important in (\ref{sim}), the expansion coefficients eventually adopting
factorial growth. For instance,  the  coefficients  of $a^5$ in (\ref{sim}) all
equal 5 approximately, while the 10th-order ones are between $5 \times 10^4$
and $9 \times 10^4$, with alternating signs.  
The functions  $W_{n}(a)$ have  divergent perturbative expansions,
resembling the expanded QCD correlation function $D$.

\begin{figure}[htb]
\begin{center} \includegraphics[width = 10cm]{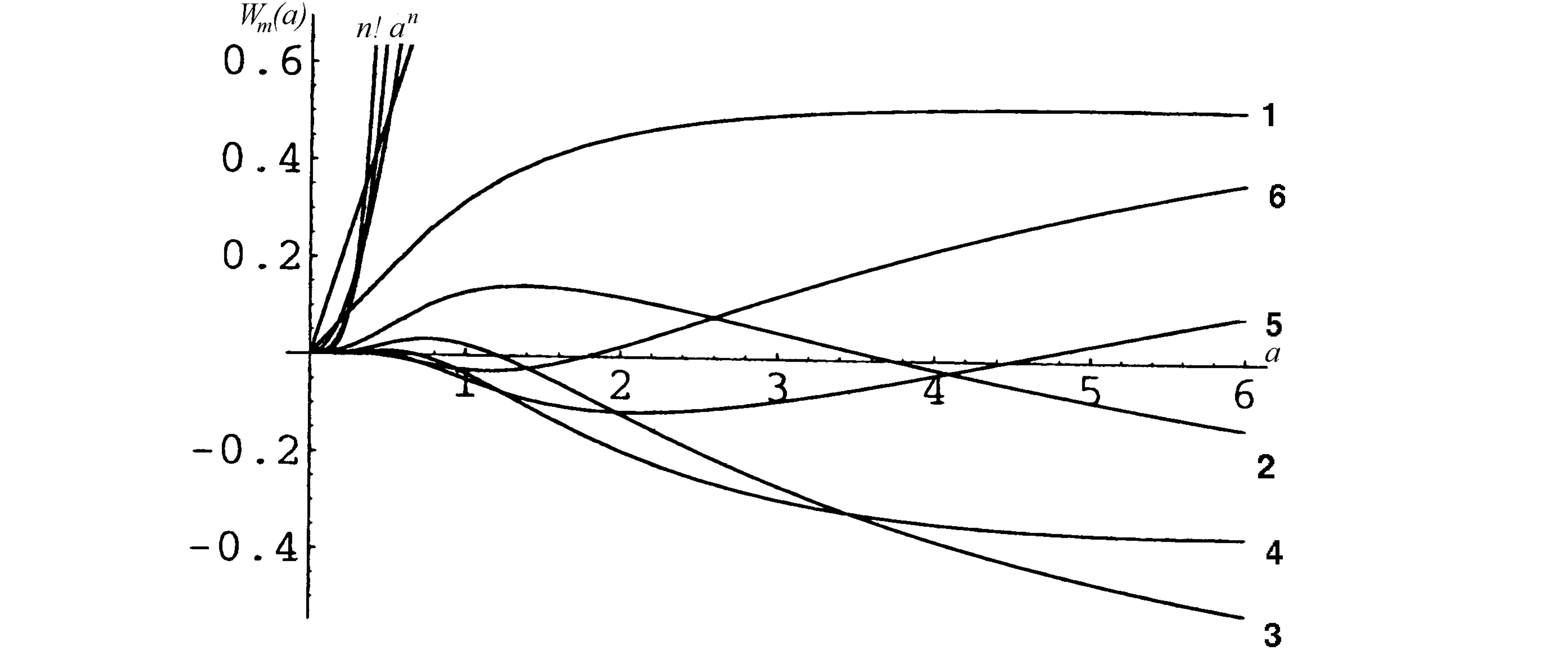}
\caption{The first six functions $W_{m}(a)$ defined with the PV
prescription (\ref{pv}), for $a$ from $0$ to $6$.  
Unlabelled are the corresponding perturbative factors $n!\, a^{n}$.\label{fig:Wa}}\end{center}
\end{figure}

Although the series (\ref{sim}) are divergent, after adopting a prescription the
functions $W_{n}(a)$ are well-defined,
and bounded in the right half plane  ${\rm Re}\, a > 0$: \be
|W_{n}(a)| \leq  \frac{ 1}{|a| } \int_{0}^{\infty}{\rm e}^{-\frac{u {\rm Re }
a}{  |a|^2}}|(\tilde w(u))^n| du  < \frac {|a|}{ {\rm Re} a} \, ,
\label{boundW} \ee since $|(\tilde w(u))^n| <1$. For $a$  real and positive
the right hand side of (\ref{boundW}) is equal to unity.  In
Fig. \ref{fig:Wa} we show, following \cite{CaFi2001},  the shape of the first functions
$W_n$, calculated with the PV prescription,  for real values of
$a$. 

Finally, an important property is the large-order behaviour of the functions $W_{n}(a)$ at large $n$.  This was investigated    \cite{CaFi2000, CaFi2001} by the technique of saddle points. Omitting the proof given in \cite{CaFi2000}, we quote  the
asymptotic  behaviour of $W_n(a)$ for $n\to \infty$:
\begin{equation}\label{Wnpsaddle} W_n(a)\approx n^{\frac{1}{ 4}} \zeta^n 
{\rm e}^{-2^{3/4} (1+i) {(n/a)^{1/2}}} + n^{\frac{1}{ 4}} (\zeta^*)^n
 {\rm e}^{-2^{3/4} (1-i) {(n/a)^{1/2}}}\,, \end{equation} where $\zeta$ was
defined below (\ref{uw}). 
This estimate  is valid in the complex $a$ plane, for  $a=|a|
{\rm e}^{{\rm i}\psi}$ with $\psi$ restricted by
\begin{equation}
|\psi|<\pi/6\,.
\label{psi}
\end{equation}

The convergence of the expansion (\ref{cW}) depends on the ratio
\begin{equation}\label{ratio}
\bigg\vert\frac{c_n W_n(a)}{ c_{n-1} W_{n-1}(a)}\bigg\vert\,.
\end{equation}
As shown in \cite{CaFi2000, CaFi2001}, if the coefficients $c_n$ satisfy the condition
 \begin{equation}\label{cnb} |c_n| < C {\rm e}^{\epsilon
n^{1/2}}\, \end{equation}
for any $\epsilon >0$, the expansion (\ref{cW}) converges for $a$ complex in the domain
\begin{equation}\label{domain}
{\rm Re}[(1\pm i) a^{-1/2}]>0\,,
\end{equation}
which is equivalent to $ |\psi|\leq \pi/2-\delta$.
Since the condition   (\ref{psi}) is more restrictive, it follows that, if the 
condition (\ref{cnb}) is satisfied,  
the series (\ref{cW}) converges in the sector defined by (\ref{psi}).

 The coefficients $c_n$ are obtained by inserting into the Taylor series (\ref{Borel}) the
expansions in powers of $w$ of the function $\tilde u(w)$ defined in (\ref{uw}). A
precise estimate of the behaviour of the $c_n$ starting from a general  form 
of the standard perturbative coefficients  $c_{n,k}$ is difficult to obtain. 
In the special case of a Borel transform with a finite number of branch-point
singularities, considered in \cite{CaFi2000, CaFi2001}, one can derive the generic
behaviour  \begin{equation}\label{cnb1} |c_n| \leq  C' n^\xi  = C' {\rm
e}^{\xi \ln n}\,,\quad \xi>0\,, \end{equation} which satisfies the convergence
condition (\ref{cnb}). Whether this bound is valid or not in general in QCD is
an open problem.

We emphasize that the 
convergence of the series (\ref{cW}) is a key argument in favour of the stepping out of the standard perturbation 
theory and the definition of a new perturbative expansion.  In the next section we shall further improve this expansion by using additional theoretical knowledge available about the expanded function. 

\section{Singularity Softening}\label{sec:soft}
In the particular case of the Adler function in massless QCD, the 
nature of the leading singularities in the Borel plane is known \cite{Mueller1992, Beneke, BBK}:
near the first branch points, $u=-1$ and $u=2$,  $B(u)$ behaves like
\begin{equation}\label{eq:gammapowers}
B(u) \sim \frac{r_1}{(1+u)^{\gamma_{1}}} \quad{\rm and} \quad B(u)  \sim \frac{r_2}{(1-u/2)^{\gamma_{2}}}, 
\end{equation}
respectively. The residues $r_1$ and $r_2$ are not known, but the exponents
$\gamma_1$ and $\gamma_2$  have known values, calculated using renormalization-group
invariance  \cite{Mueller1992, BBK, Beneke, BeJa}:
\begin{equation}\label{eq:gamma12}
\gamma_1 = 1.21,    \quad\quad   \gamma_2 = 2.58 \,. 
\end{equation}

The expansion (\ref{Bw}) takes into account only the position 
of the renormalons in the Borel plane. If a sufficient number of expansion 
coefficients were known, (\ref{Bw}) would be expected to describe also the 
character, strength, etc., of the singularities as well. Since, however, only a few
 perturbative coefficients are at present explicitly available, 
one cannot expect that the expansion of the type (\ref{Bw}) might be 
able to give a satisfactory approximation of $B(u)$ near its first singularities. It is better than 
(\ref{Borel}), which has no singularities in any finite-order approximation. But, although the position of the first singularities is correctly implemented by (\ref{Bw}), their nature cannot be captured by a few number of terms in the expansion.   

An explicit account for the leading singularities (\ref{eq:gammapowers}) would
therefore be helpful to further improve the convergence. This can be done by multiplying
$B(u)$ with suitable factors that vanish at $u=-1$ and $u=2$ and compensate  the dominant singularities.
The subsequent expansion of the product in powers  of a conformal mapping variable is expected to 
converge better. This procedure is known as  "singularity softening" \cite{SoSu, CaFi1998, CaFi2009, CaFi_RJP, CaFi2011, CaFi_Manchester}.  

In contrast with the optimal conformal mapping, singularity softening is not unique. 
The singularities are 
present in $B(u)$, but we do not know their actual form, except for the behavior (\ref{eq:gammapowers}) near 
the corresponding branch-points. A possibility is to multiply $B(u)$ by
simple factors like $(1+u)^{\gamma_1} (1-u/2)^{\gamma_2}$   \cite{SoSu,CaFi1998}.
In \cite{CaFi2009}, the alternative softening factors $(1+w)^{2 \gamma_1} (1-w)^{2 \gamma_2}$ were adopted, 
where $w=\wt w(u)$ is the optimal mapping (\ref{eq:w}). The product of $B(u)$ with these factors was afterwards expanded in powers of the same variable $w$. 

In fact, some generalizations of this expansion can be constructed. We note that the product of $B(u)$ with softening factors is expected to contain milder singularities,  which vanish
instead of becoming infinite at  $u=-1$ and $u=2$  (in very peculiar cases the singularities may disappear altogether, but this situation is very unlikely).  The effect of a mild singularity in a function is not visible at low orders  in its series expansions, and is  expected to appear only at large orders.  Therefore, we can ignore their effects, expanding  the product 
in powers of variables that account only for the next branch points of $B(u)$. In the case of the Adler function, these singularities are placed at $u=3,\,4$, etc., on the positive axis, and at  $u=-2,\,-3$, etc., on the negative axis.

\begin{figure}[!ht]
\begin{center}
\includegraphics[width=5.5cm]{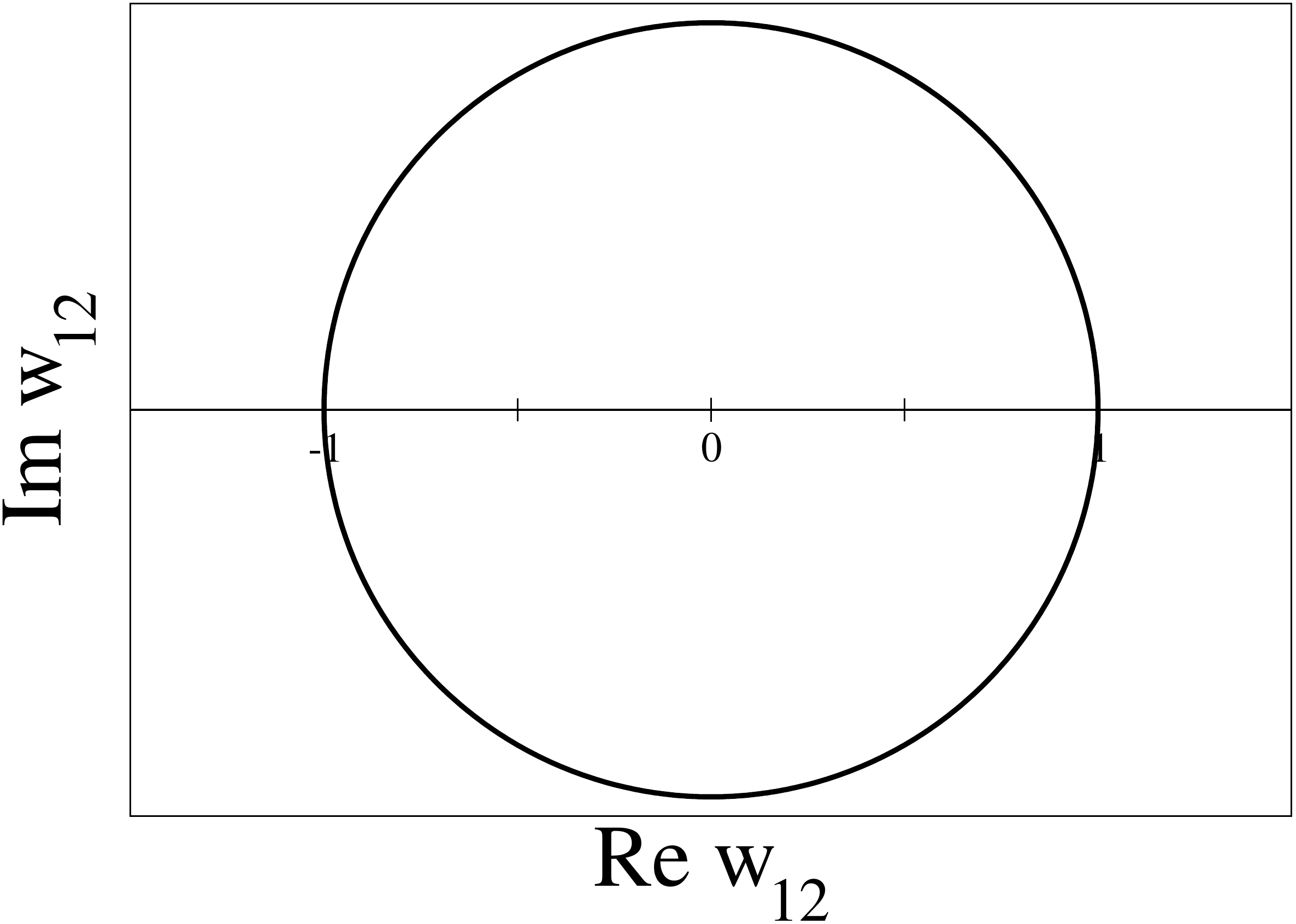}\hspace{0.3cm}
\includegraphics[width=5.5cm]{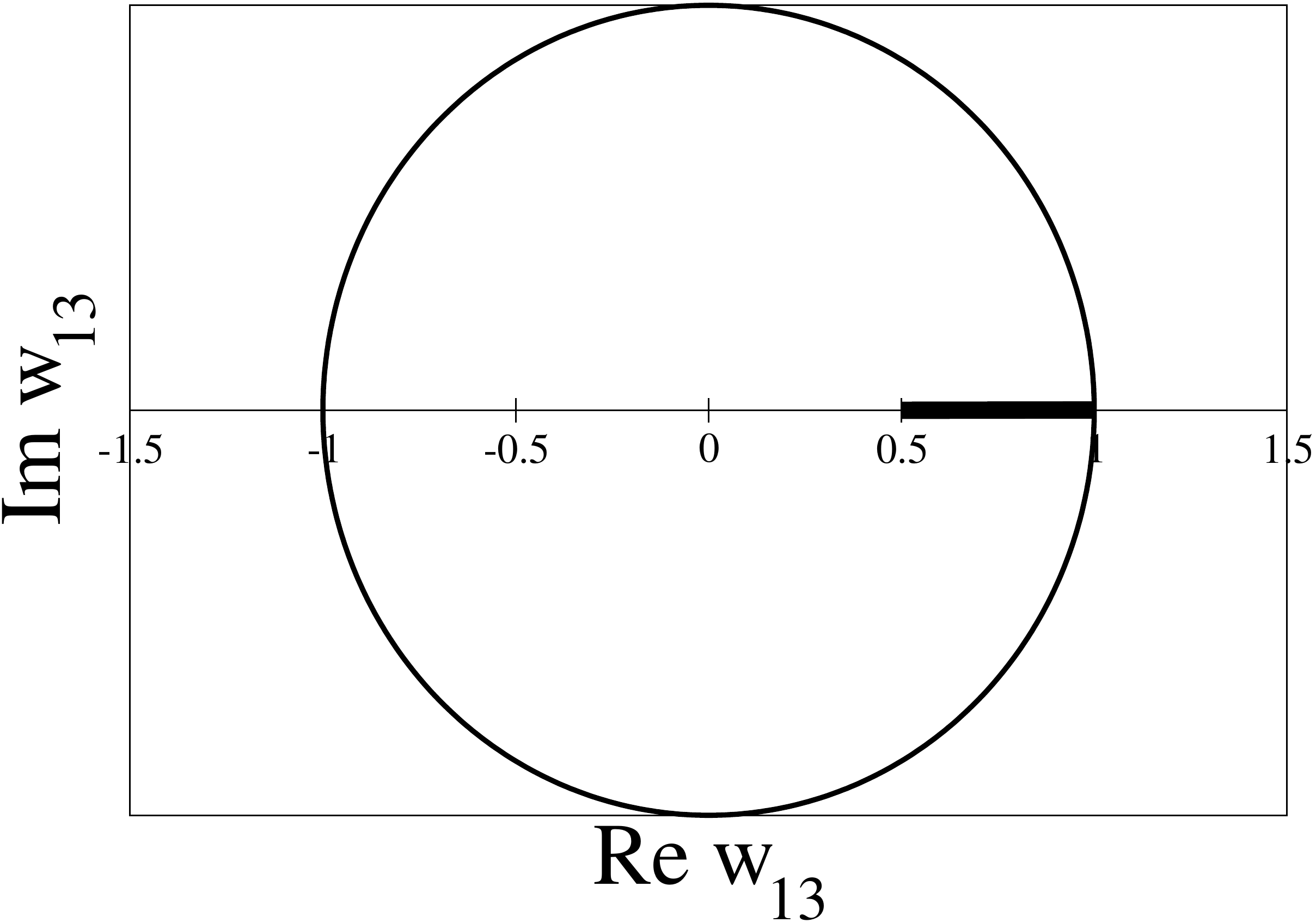}\\\includegraphics[width=5.5cm]{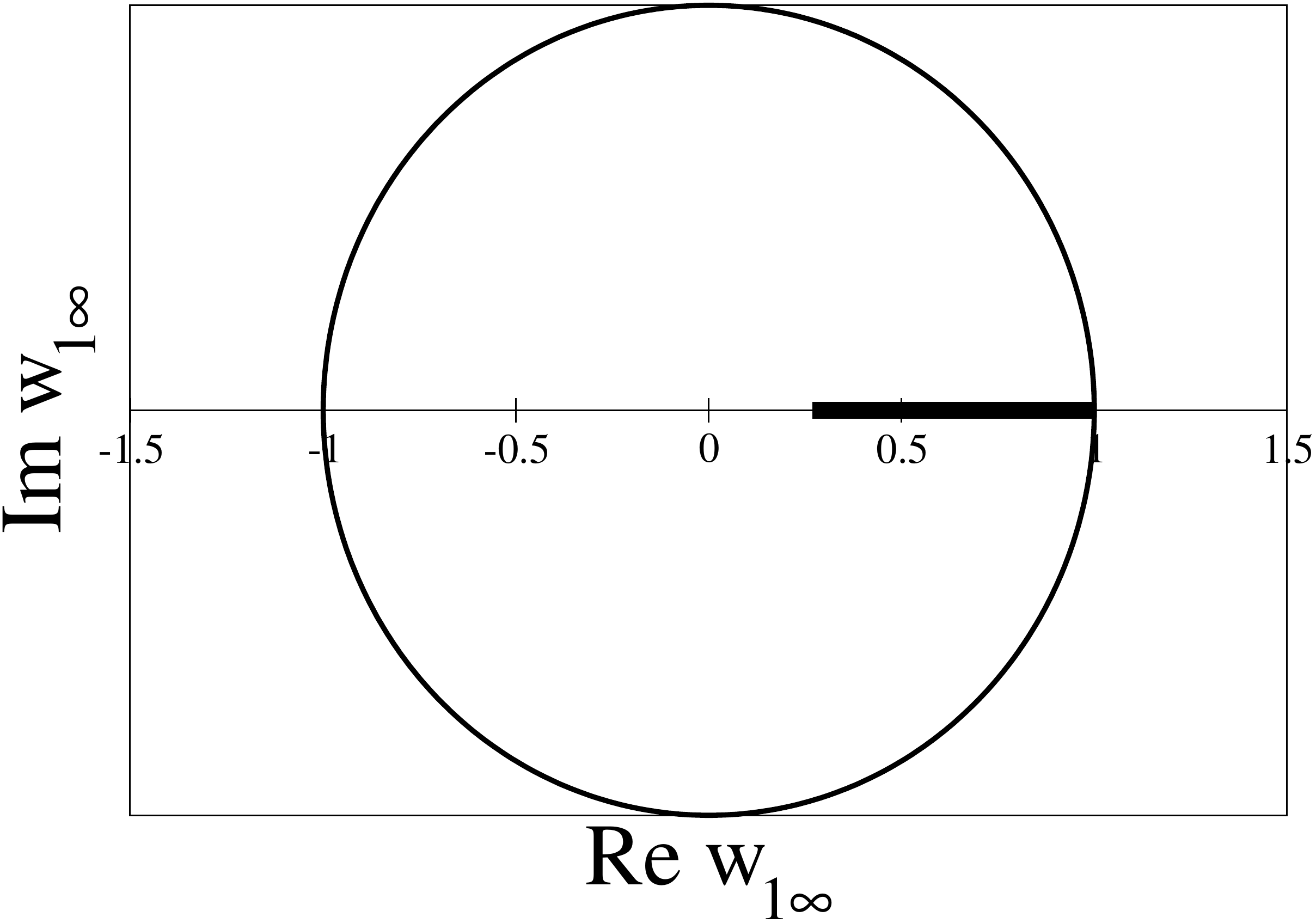}\hspace{0.3cm}\includegraphics[width=5.5cm]{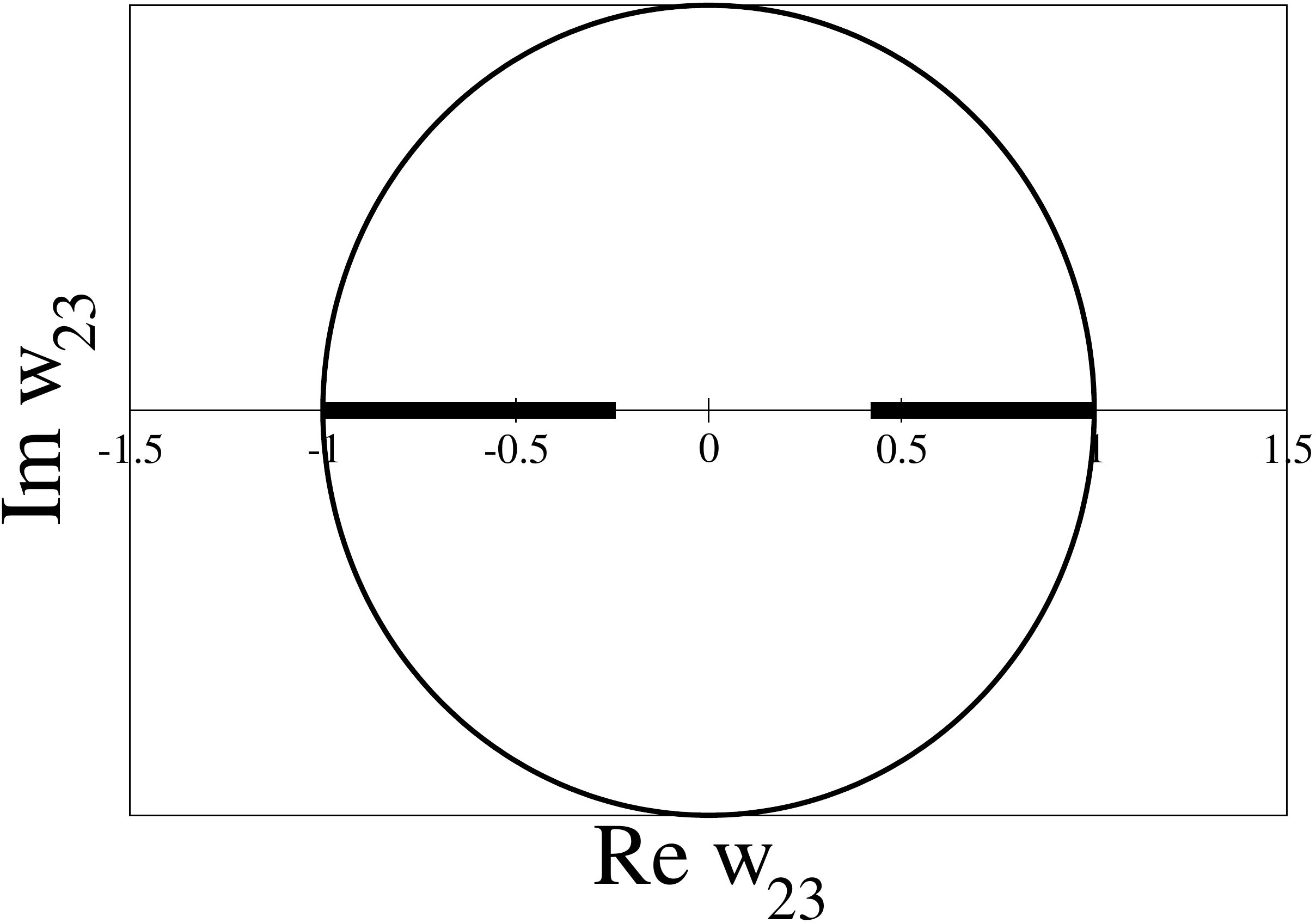}
\end{center}
\caption{\label{fig:wjk}   The unit disks $|w_{jk}|<1$ on which the conformal mapping defined in (\ref{eq:wjk}) maps the cut $u$-plane, for several values of $j$ and $k$ \cite{CaFi2011}. In the last three figures, the thick lines indicate the residual cuts inside the unit disk.  }
\end{figure}

It is  useful then to define the generic functions \cite{CaFi2011}
\be\label{eq:wjk}
\wt w_{jk}(u)=\frac{\sqrt{1+u/j}-\sqrt{1-u/k}}{\sqrt{1+u/j}+\sqrt{1-u/k}},\ee
 which conformally map the $u$ plane cut along $ u\le -j$ and $u\ge k$ to the disk $|w_{jk}|<1$ in the plane $w_{jk}\equiv \wt w_{jk}(u)$.
For $j= 1$, $k= 2$, we obtain the optimal mapping (\ref{eq:w}).  In the following, we shall consider also the variables  $w_{13}$, $w_{1\infty}$ and  $w_{23}$, for which the corresponding unit disks $|w_{jk}|<1$ are shown  in Fig. \ref{fig:wjk}.  The conformal mapping $w_{1\infty}$ coincides actually with the mapping  (\ref{v}) suggested in \cite{Mueller1992} and the mapping  $w_{13}$ was investigated  also in \cite{CvLe}.  As seen  in Fig.  \ref{fig:wjk}, the last three mappings leave inside the unit circle parts of the real axis of the $u$ plane which contain some singularities. As a consequence, the expansions based on these variables will converge in a smaller domain and their convergence rates  will be, in principle, worse than that of the optimal mapping $w_{12}$.

According to the above discussion, we shall expand in powers of  $w_{jk}$ the product of $B(u)$ with suitable softening factors. Specifically, we consider the expansions  \cite{CaFi2011}
\be \label{eq:prod} S_{jk}(u) B(u) = \sum_{n\ge 0} c_n^{jk}  (\wt w_{jk}(u))^n,
\ee
where  $S_{jk}(u)$ must ``soften" in principle all the singularities of $B(u)$ at  $-j\leq u<0 $ and $0<u \leq k$.
 
A systematic application of this idea to the singularities of $B(u)$ requires the knowledge of the nature of the branch-points, which at present is available only for the leading singularities at $u=-1$ and $u=2$.  
 Therefore, we shall limit ourselves to compensating factors that vanish at these points.
Numerically, it is convenient to choose the factor $S_{jk}$ as a simple expression with a 
rapidly converging expansion in powers of $w_{jk}$, thus ensuring  a good convergence of the product (\ref{eq:prod}).
A suitable choice is \cite{CaFi2011}:
 \be\label{eq:Sjk}
 S_{jk}(u)=\left(\!1-\frac{\wt w_{jk}(u)}{\wt w_{jk}(-1)}\!\right)^{\!\!\gamma^{(j)}_1} \!\!\left(\!1-\frac{\wt w_{jk}(u)}{\wt w_{jk}(2)}\!\right)^{\!\!\gamma^{(k)}_2}. \ee
The exponents $\gamma_1^{(j)}=\gamma_1 (1+\delta_{j1})$ and 
 $\gamma_2^{(k)}= \gamma_2 (1+\delta_{k2})$, where $\delta_{ij}$ is the Kronecker delta, are taken such as to reproduce the nature of the first branch-points of $B(u)$, given in (\ref{eq:gammapowers}). In particular, for the optimal case $j=1$, $k=2$ we obtain from (\ref{eq:Sjk}) the factor $(1+w)^{2 \gamma_1}(1-w)^{2 \gamma_2}$, with $w=\wt w(u)$ defined in (\ref{eq:w}).

Strictly speaking, for a fixed pair ($j, k$) the expansion (\ref{eq:prod})  converges only on the disk $|w_{jk}|< \min [|\wt w_{jk}(-1)|,\, |\wt w_{jk}(2)|]$. For the optimal choice $j=1, k=2$, the expansion converges in the whole unit disk $|w_{12}|<1$, {\em i.e.} in the whole $u$ plane except for the cuts along the real axis for $u\ge 2$ and $u\le -1$.  For other mappings, the convergence disk is limited by the beginning of the cuts shown in Fig. \ref{fig:wjk}. In particular,  if $j=1$ and $k>2$ the expansions (\ref{eq:prod})  diverge for real $u$ greater than 2,  while for the conformal mappings with $j>1$,  the expansions start to diverge for $u$ greater than one, due to the singularity at $u=-1$ present inside the circle (as in the last case shown in Fig. \ref{fig:wjk}).  However, for the product $S_{jk}(u) B(u)$ these singularities are mild.

The expansion (\ref{eq:prod})  enters the Laplace-Borel integral (\ref{PV}) where,  for values of  $a$  of physical interest, the contribution of high values of $u$ is suppressed. In particular, if $a$ is not very large, the region $u>2$ brings a small contribution to the integral, so signs of divergence in the case of the variables  $w_{13}$ and  $w_{1\infty}$ are expected to occur only at very large orders $N$. On the other hand, for the variable $w_{23}$, it is natural to expect signs of divergence  at lower values of $N$, since the series (\ref{eq:prod}) does not converge for $u>1$ . 

By combining the expansion (\ref{eq:prod}) with the definition (\ref{PV}), we are led to the general class of perturbative expansions 
\be\label{eq:Djk}  D(s) = \sum\limits_{n=0}^\infty c_n^{(jk)} \, W^{(jk)}_n(a),\ee
in terms of the expansion functions
\be\label{eq:Wnjk} W^{(jk)}_n(a)=\frac{1}{a} {\rm PV}\int\limits_0^\infty\!e^{-\frac{u}{a}} \,\frac{(\wt w_{jk}(u))^n}{S_{jk}(u)}\, d u.\ee

The properties of these expansions are similar to those of the simpler functions $W_n$ presented in the previous section. In sections \ref{sec:models} and \ref{sec:tau} we shall discuss the application of these expansions both to mathematical toy models and for the extraction of the strong coupling from hadronic $\tau$ decays. Before turning to this, we need to make a brief digression by analyzing another source of ambiguity  of perturbative QCD at finite orders, which is the subject of the next section.

\section{Renormalization-Group Summations}\label{ref:RG}
The Adler function is by definition renormalization-group invariant. However, this property is no longer valid for its perturbative expansions truncated at finite orders, which depend both on renormalization scheme and scale. In our presentation we shall work in a fixed scheme ($\overline{\rm MS}$) and concentrate on the dependence on  scale. For convenience, in what follows we shall write the Adler function as 
\be\label{whD}
D(s)=1+\wh D(s),
\ee
where the first term is the parton model result, and consider only the nontrivial contribution $\wh D(s)$, whose perturbative expansion is given in (\ref{Ds}) and traditionally called  ``fixed-order perturbation theory'' (FOPT).  Using (\ref{Ds}), we write
\begin{equation}
\label{eq:FOPT}
\wh D_{\rm FOPT}(s) = \sum\limits_{n\ge 1} \,\left(\frac{\alpha_s(\mu^2)}{\pi}\right)^n\,
\sum\limits_{k=1}^{n} k\, c_{n,k}\, (\ln (-s/\mu^2))^{k-1} \,.
\end{equation}
The renormalization-group improved expansion (\ref{DsRG}), which we used so far in our discussion, is also called, for reasons that will become clear in the next sections, ``contour-improved  perturbation theory'' (CIPT). Thus, using (\ref{DsRG}) we have:
\be\label{eq:CIPT}
\wh D_{\rm CIPT}(s) = \sum\limits_{n\ge 1} \, c_{n,1} \left(\frac{\alpha_s(-s)}{\pi}\right)^n\,. 
\ee

We shall consider also another approach, proposed in \cite{Ahmady1,Ahmady2}, which generalizes the summation of leading logarithms by summing all the terms available from renormalization-group  invariance.  This formulation of perturbation theory, applied to the Adler function in \cite{Abbas:2012py, Abbas:2012fi}, is referred to as ``renormalization-group-summed perturbation theory'' (RGSPT). For our purpose, it is useful to note that the expansion of the Adler function can be written as \cite{Abbas:2012fi}  
\be\label{eq:RGSPT}\wh D_{\rm RGSPT}(s) = \sum_{n\ge 1} \,(\wt a_s(-s))^n \left[c_{n,1}+ \sum_{j=1}^{n-1} c_{j,1} d_{n,j}(y)\right], 
 \ee
where $\wt a_s(-s)$ is the solution  of the RG equation (\ref{eq:rge})  to one loop, given by (\ref{a1loop}), and the functions $d_{n,j}(y)$ have analytically closed forms depending only on  the variable $y\equiv 1+\beta_0 a_s(\mu^2) \ln(-s/\mu^2)$, where $\beta_0$ is the first coefficient of the $\beta$ function, given in (\ref{betaj}). The explicit expressions of these functions for $n\le 10$ can be found in \cite{Abbas:2012py, Abbas:2012fi}.

At finite truncation orders, the difference between the predictions of the above three versions of perturbation theory produces an unavoidable theoretical ambiguity, which affects the extraction of the QCD parameters from experimental measurements. As noticed in \cite{BeJa}, the corresponding theoretical error of the strong coupling $\alpha_s$ at the scale $\mu^2=m_\tau^2$, determined from the hadronic decays of the $\tau$ lepton, turned out to  increase instead of decreasing when higher-order loop calculations of the Adler function were  included. This surprising result  has generated many debates and controversial opinions on how to handle it have been formulated  \cite{BeJa,  Pich_Muenchen, DeMa, BBJ}. It shows actually that the uncertainty due to renormalization-group summations is  correlated to the behaviour of the higher-order coefficients and the divergency of the series. Both effects are relevant for predictions
at the $m_\tau$ scale, where the coupling $\alpha_s$ is rather large. It would be interesting therefore to define improved expansions, based on the ideas of conformal mappings and singularity softening, also for the FOPT and RGSPT series defined above.

It is convenient to define the Borel transform $\wh B_{\rm CIPT}(u)$ of the expansion $\wh D_{\rm CITP}(s)$ by the somewhat different expansion: 
\be\label{eq:Bhat}
\wh B_{\rm CIPT}(u)= \sum_{n=0}^\infty \wh b_n u^n,\quad\quad \wh b_n= \frac{c_{n+1,1}}{\beta_0^n \,n!}\,,
\ee
which implies the Laplace-Borel integral representation 
\be\label{eq:Dhat}
\wh D_{\rm CITP}(s)=\frac{1}{\beta_0}\,{\rm PV} \,\int\limits_0^\infty  
\exp{\left(\frac{-u}{\beta_0 a_s(-s)}\right)} \, \wh B_{\rm CIPT}(u)\, d u\,.
\ee
By analogy with (\ref{eq:Djk}) and (\ref{eq:Wnjk}), we can write the improved perturbative CIPT expansion of the Adler function:
\be\label{eq:Dhatjk}  \wh D_{\rm CITP}(s) = \sum\limits_{n=0}^\infty c_{n, {\rm CITP}}^{(jk)} \,\wh W^{(jk)}_{n, {\rm CITP}}(s),\ee
where the expansion functions have the expression
\be\label{eq:Whatnjk}\wh W^{(jk)}_{n, {\rm CITP}}(s)=\frac{1}{\beta_0 } {\rm PV}\int\limits_0^\infty\! e^{-\frac{u}{\beta_0 a_s(-s)  }} \,\frac{(\wt w_{jk}(u))^n}{S_{jk}(u)}\, d u\,,\ee
and the coefficients $c_{n,{\rm CITP}}^{(jk)}$ are obtained from Eqs. (\ref{eq:wjk}), (\ref{eq:Sjk}) and (\ref{eq:Bhat}). 

To emphasize the fact that the expansion functions (\ref{eq:Whatnjk}) are no longer powers of the coupling $a_s$, the expansion (\ref{eq:Dhatjk}) is sometimes called ``non-power perturbation theory'' (NPPT) \cite{CaFi2011,Abbas:2012fi}.

Similar non-power expansions  can be defined also for the FOPT and RGSPT versions of perturbation theory. In these cases, the Borel transforms $\wh B_{\rm FOPT}(u,s)$ and $\wh B_{\rm RGSPT}(u, s)$, respectively, defined  starting from the expansions (\ref{eq:FOPT}) and (\ref{eq:RGSPT}), depend also on the variable $s$.  However, as discussed in \cite{Abbas:2012fi}, the position and nature of the  leading singularities in the $u$ plane of these Borel transforms  are identical to  those of $\wh B_{\rm CIPT}(u)$. This result follows from a 
general argument by Mueller \cite{Mueller1985}, which states that the dominant singularities of the Borel
transform are determined from the behaviour of the correlators in the limit of small
coupling, when the three different couplings relevant for the above expansions, namely $a_s (-s)$, $a_s (m_\tau^2)$ and $\tilde a_s(-s)$, are close to each other.  Therefore, the optimal conformal mapping $\tilde w(u)$ defined in (\ref{eq:w}), as well as the more general mappings (\ref{eq:wjk}) and softening factors (\ref{eq:Sjk}) defined above, remain the same in the case of FOPT and RGSPT.  The corresponding improved  expansions, similar to Eqs. (\ref{eq:Dhat})-(\ref{eq:Whatnjk}),  can be found in Ref. \cite{Abbas:2012fi} and are not repeated here.

\section{Toy Models}\label{sec:models}

The convergence properties of the expansions discussed  above  have been tested through toy theoretical models which predict the higher-order coefficients of the Adler function, $c_{n,1}$ for $n>4$.  In these models, the Borel transform is expressed in terms of a few dominant singularities in the Borel plane.

In a first theoretical model, proposed in \cite{BeJa} and discussed in many papers as a reference model,  the  Adler function $\wh D(s)$ is defined as the {\rm PV}-regulated Laplace-Borel integral (\ref{eq:Dhat}), 
where the Borel transform $\wh B(u) \equiv \wh B_{\rm ref}(u)$ is expressed in terms of a few ultraviolet (UV) and infrared (IR) renormalons, and a regular, polynomial part:
\be\label{eq:BBJ}
\frac{\wh B_{\rm ref}(u)}{\pi}=B_1^{\rm UV}(u) +  B_2^{\rm IR}(u) + B_3^{\rm IR}(u) +
d_0^{\rm PO} + d_1^{\rm PO} u, 
\ee
where  the renormalons are parametrized as \cite{BeJa}
\bea\label{eq:BIRUV}
B_p^{\rm IR}(u)= 
\frac{d_p^{\rm IR}}{(p-u)^{\gamma_p}}\,
\Big[\, 1 + \wt b_1 (p-u) + \ldots \,\Big], \nonumber \\
B_p^{\rm UV}(u)=\frac{d_p^{\rm UV}}{(p+u)^{\bar\gamma_p}}\,
\Big[\, 1 + \bar b_1 (p+u)  +\ldots \,\Big].
\eea

The free parameters of the model are determined such that they reproduce  the known perturbative coefficients $c_{n,1}$ for $n\le 4$ given in (\ref{cn1}), and the estimate $c_{5,1}=283$ for the next coefficient. Their numerical values are \cite{BeJa}:
{\be\label{eq:dBJ}
d_0^{\rm PO}=0.781, ~~~
d_1^{\rm PO}=7.66\times 10^{-3},~~
 d_2^{\rm IR}=3.16,~~ d_3^{\rm IR}=-13.5,~~ d_1^{\rm UV}=  - 1.56 \times 10^{-2}. 
\ee}
After specifying the parameters, all the higher-order coefficients $c_{n,1}$ can be predicted and they exhibit a factorial growth. Their numerical values  up to $n=18$ are listed in Refs. \cite{BeJa,CaFi2009}.

From (\ref{eq:dBJ}) one can see that this model has a relatively large residue $d_2^{\rm IR}$ of the first IR renormalon at $u=2$.
However, a smaller residue $d_2^{\rm IR}$ of the first IR renormalon is not excluded for the physical function.  Models attempting to simulate this situation have been investigated in several papers. For instance, an extreme alternative model,  with no singularity at all at $u=2$ and an additional singularity at $u=4$, is defined by choosing the Borel transform  as:
\be\label{eq:altBBJ}
\frac{\wh B_{\rm alt}(u)}{\pi}=B_1^{\rm UV}(u) +  B_3^{\rm IR}(u) + B_4^{\rm IR}(u)  
+ d_0^{\rm PO} + d_1^{\rm PO} u.
\ee
The five parameters,  found by matching the coefficients $c_{n,1}$ for 
$n\le 5$, are: 
{\be\label{eq:altdBJ}
d_0^{\rm PO}=2.15, ~~ d_1^{\rm PO}=4.01 \times 10^{-1},~~
d_3^{\rm IR}=66.18, ~~ d_4^{\rm IR}=-289.71, ~~d_1^{\rm UV}=-5.21 \times
10^{-3}.
\ee}
Several intermediate models, including the first IR renormalon at $u=2$ and a prescribed residue smaller (or larger) than the value in (\ref{eq:dBJ}) have been also investigated in Refs. \cite{ CaFi2011,  Abbas:2012fi, Pich_Muenchen, DeMa, BBJ, Boito2013, Abbas:2013usa, Caprini:2013bba}.

Before applying  the perturbative expansions to the toy models, we recall that perturbative QCD is not directly applicable at low energies on the timelike axis \cite{PQW}: indeed, the correlation functions and the scattering amplitudes exhibit in this region hadronic thresholds implied by unitarity, which cannot be  described in terms of free quarks and gluons. Moreover, the perturbation series becomes useless, since the running coupling $\alpha_s(-s)$ is very large at low $s>0$.

 Therefore, we shall test the various expansions by calculating the values of the Adler function $\wh D(s)$ in the complex $s$ plane, outside the real positive axis. Having in mind the physical applications, in particular to the hadronic $\tau$ decays, we shall calculate the function along the circle $|s|=m_\tau^2$, {\em i.e.}  for  $s= m_\tau^2 \exp (i \varphi)$, with $0\leq \varphi\le 2\pi$. Actually, using the Schwarz reflection property $\Pi(s^*)=\Pi^*(s)$ satisfied by the polarization function, it is enough to consider only the range   $0\leq \varphi\le \pi$. 

The exact function $\wh D(s)$ is obtained by inserting in  (\ref{eq:Dhat}) the Borel transform  (\ref{eq:BBJ}) or (\ref{eq:altBBJ}) and using in the exponent the solution $\alpha_s(-s)$ of the renormalization-group equation (\ref{eq:rge}), found numerically in an iterative way along the circle, starting from a given initial  value at  $s=- m_\tau^2$. In all the calculations we have used for convenience the value $\alpha_s(m_\tau^2)= 0.34$.

The new non-power perturbative expansions  are constructed by truncating the standard series at a definite value N and passing to the new expansions by using the algorithm presented in the previous section. For a fixed N, the new expansions reproduce the coefficients $c_{n,1}$ with $n\leq $N. In the CIPT version the $s$-dependent coupling $\alpha_s(-s)$ is calculated along the circle by integrating numerically the RG equation, as explained above. The FOPT version involves only the coupling  $\alpha_s(m_\tau^2)$, but contains an additional $s$ dependence in the expansion coefficients. The RGS version involves the one-loop coupling $\tilde\alpha_s(-s)$ and a residual $s$ dependence in the coefficients.

\begin{figure}[htb]\vspace{0.5cm}
\begin{center} \includegraphics[width =6cm]{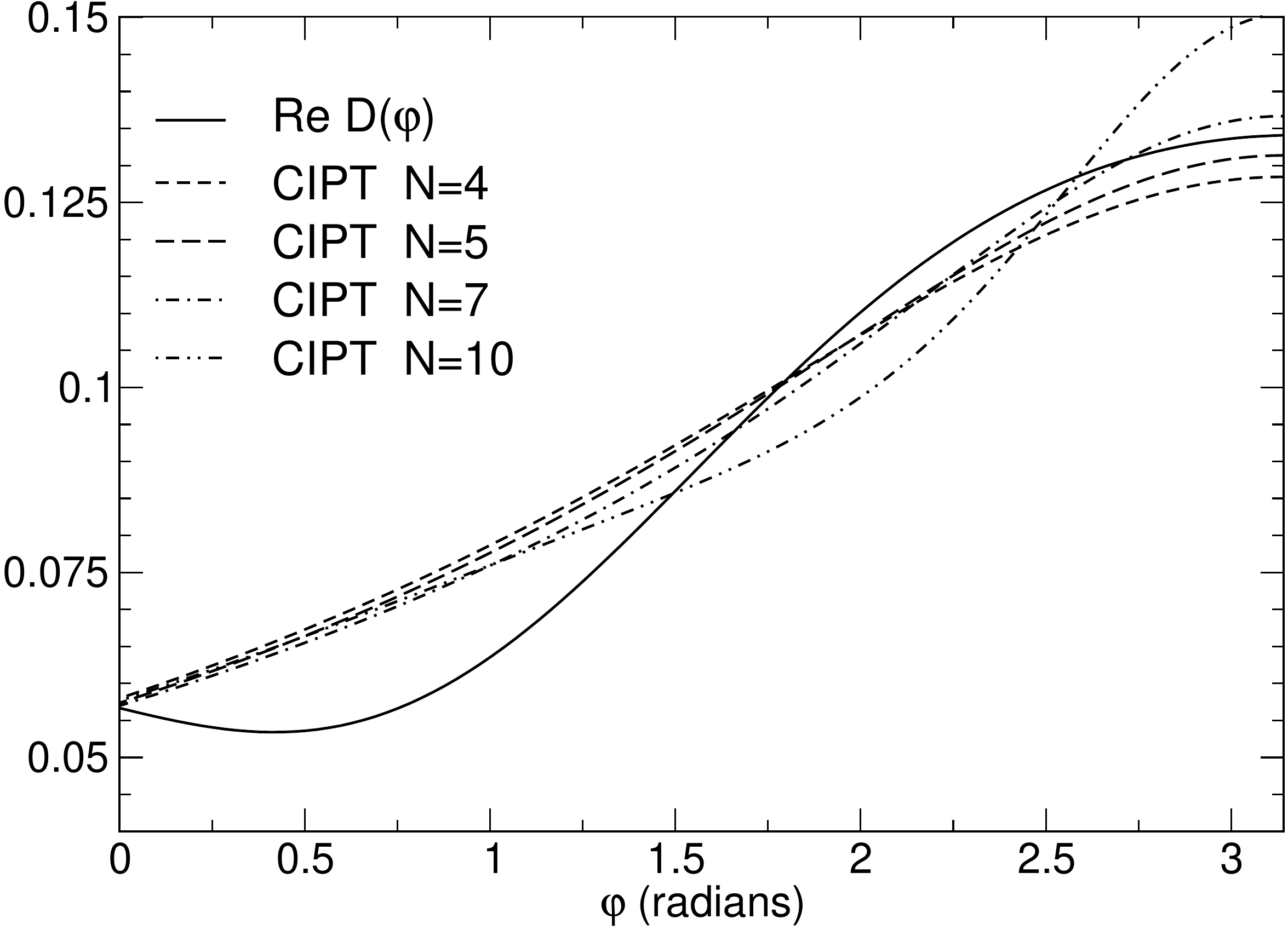}\hspace{1.5cm}\includegraphics[width =6cm]{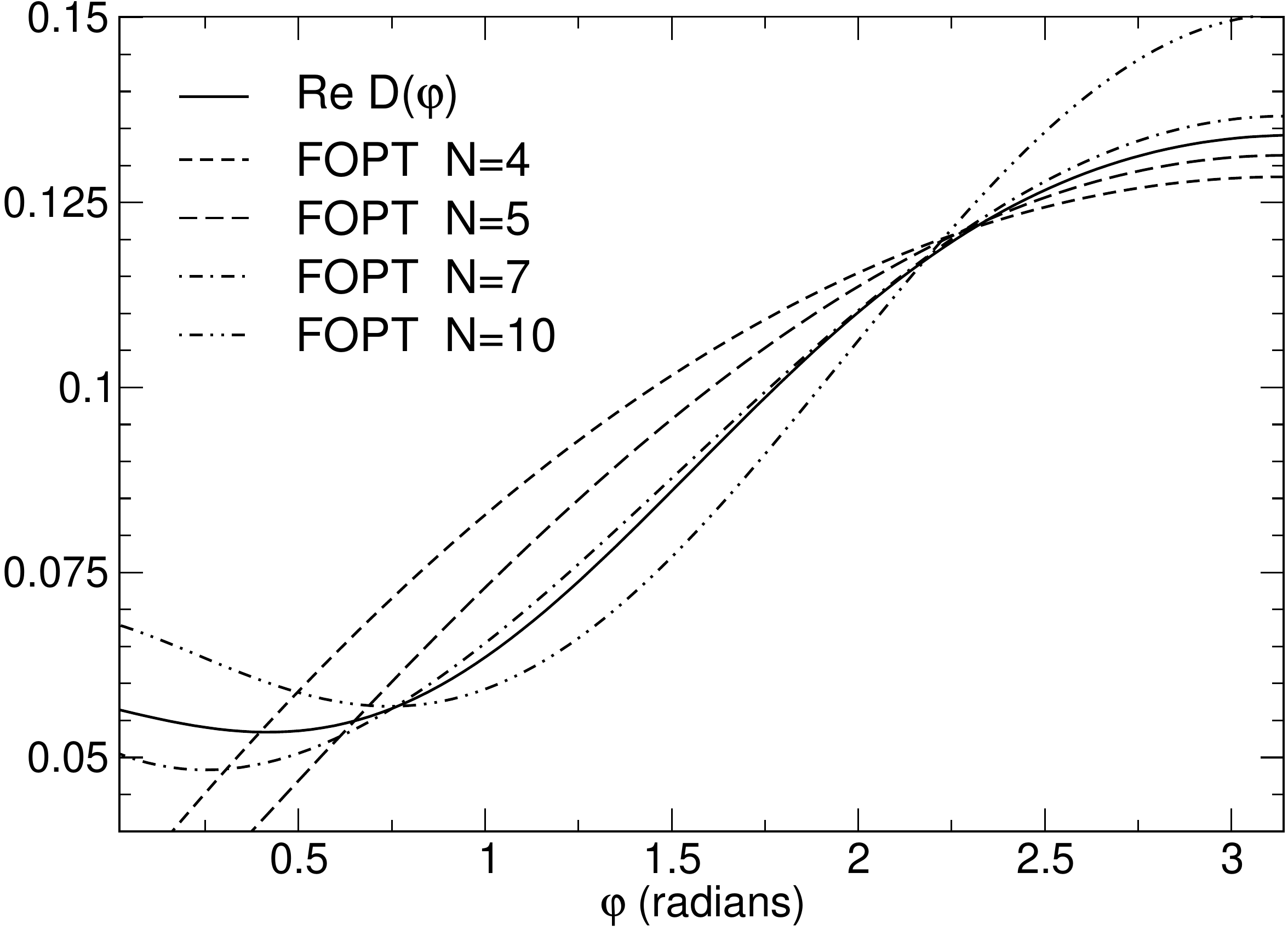}
\caption{ Real part of the Adler function in the model (\ref{eq:BBJ}) along the circle $s=m_\tau^2 e^{i\varphi}$ for $0\leq \varphi\le \pi$  (solid line)  and its perturbative values calculated with the standard  CIPT (left panel) and FOPT (right panel) expansions truncated after N terms \cite{BeJa, CaFi2009}.  For higher perturbative orders N the expansions show big oscillations  and are not shown. \label{fig:DRCIFO}}\end{center}
\end{figure}

 We start by illustrating the properties of the standard expansions in  Fig. \ref{fig:DRCIFO}, where we  show the real part of the  Adler function for the model (\ref{eq:BBJ}) and its approximants calculated with the standard CIPT and FOPT expansions along the circle $|s|=m_\tau^2$. One can see that the description is not very good: neither CIPT nor FOPT succeed in approximating with precision the exact function, and the approximation becomes worse when the truncation order N is increased.  In particular, FOPT fails to approach the exact function neither near the timelike axis, which corresponds to $\varphi=0$, nor near the spacelike axis,  which corresponds to
$\varphi=\pi$. 

\begin{figure}[htb]\vspace{0.5cm}
\begin{center} \includegraphics[width =6cm]{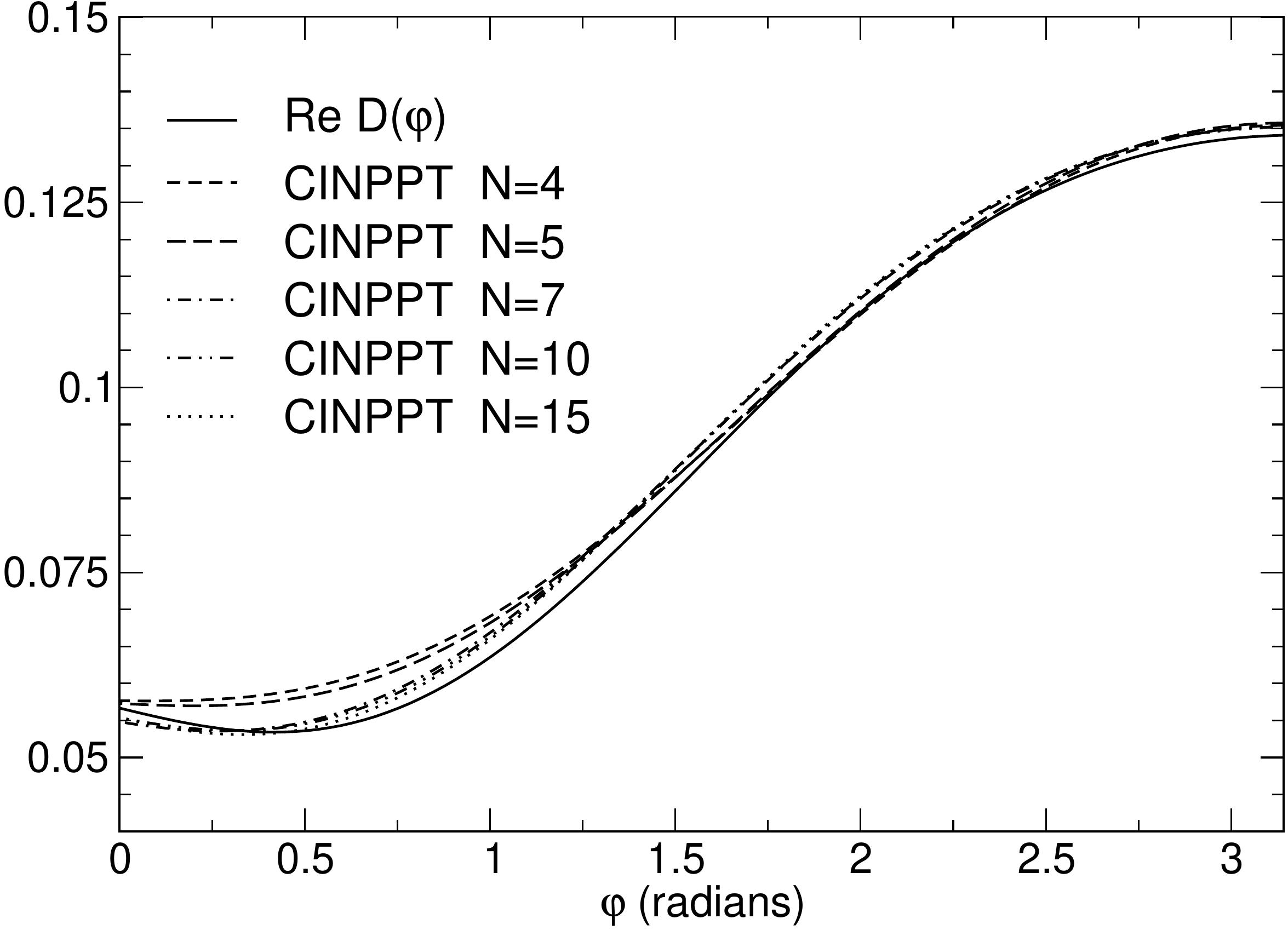}\hspace{1.5cm}\includegraphics[width =6cm]{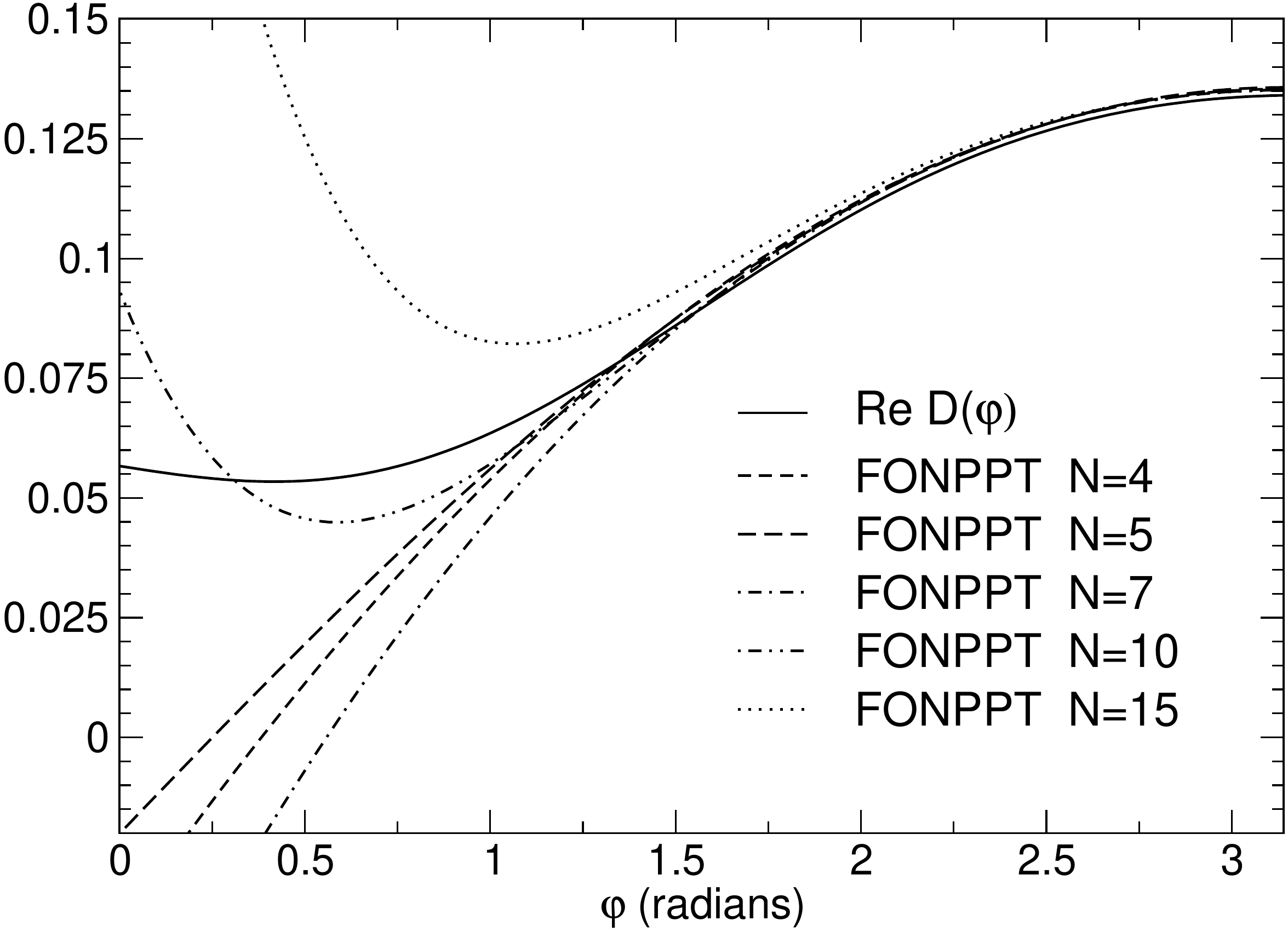}
\caption{ As in Fig. \ref{fig:DRCIFO} using the non-power perturbation theory (NPPT) defined in (\ref{eq:Dhat}), with the optimal conformal mapping $w_{12}$.  \label{fig:DRCIFOw}}\end{center}
\end{figure}

For comparison, we illustrate in Fig. \ref{fig:DRCIFOw} the properties of the new, non-power expansion (\ref{eq:Dhat}), in the CI and FO versions. The optimal conformal mapping $w_{12}$ has been used in the calculations. The left panel proves the very good approximation achieved by the perturbative expansions improved by both renormalization-group and the analytic continuation in the Borel-plane, up to high perturbative orders. As for the FO non-power expansions, they provide a very good approximation for points near the spacelike axis ($\varphi$ close to $\pi$). Near the timelike axis, the description is worse due to the large imaginary parts in the logarithm $\ln(-s/m_\tau^2)$ and its powers, appearing in the coefficients. This poor convergence near the timelike axis is an intrinsic feature of the FO expansions, which is manifest also in the case of the series improved by conformal mappings of the Borel plane.

\begin{figure}[htb]\vspace{0.5cm}
\begin{center} \includegraphics[width =6cm]{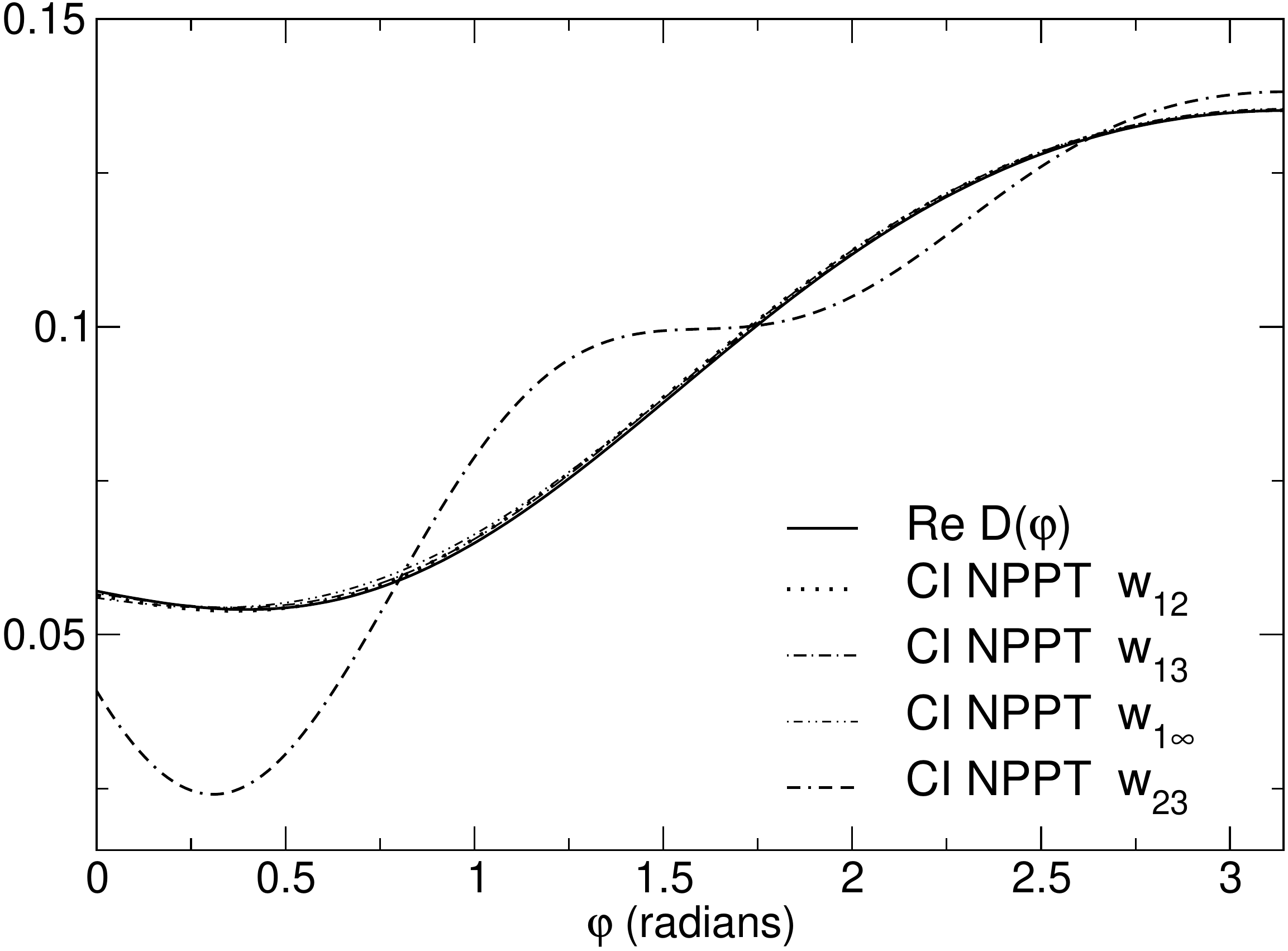}\hspace{1.5cm}\includegraphics[width =6cm]{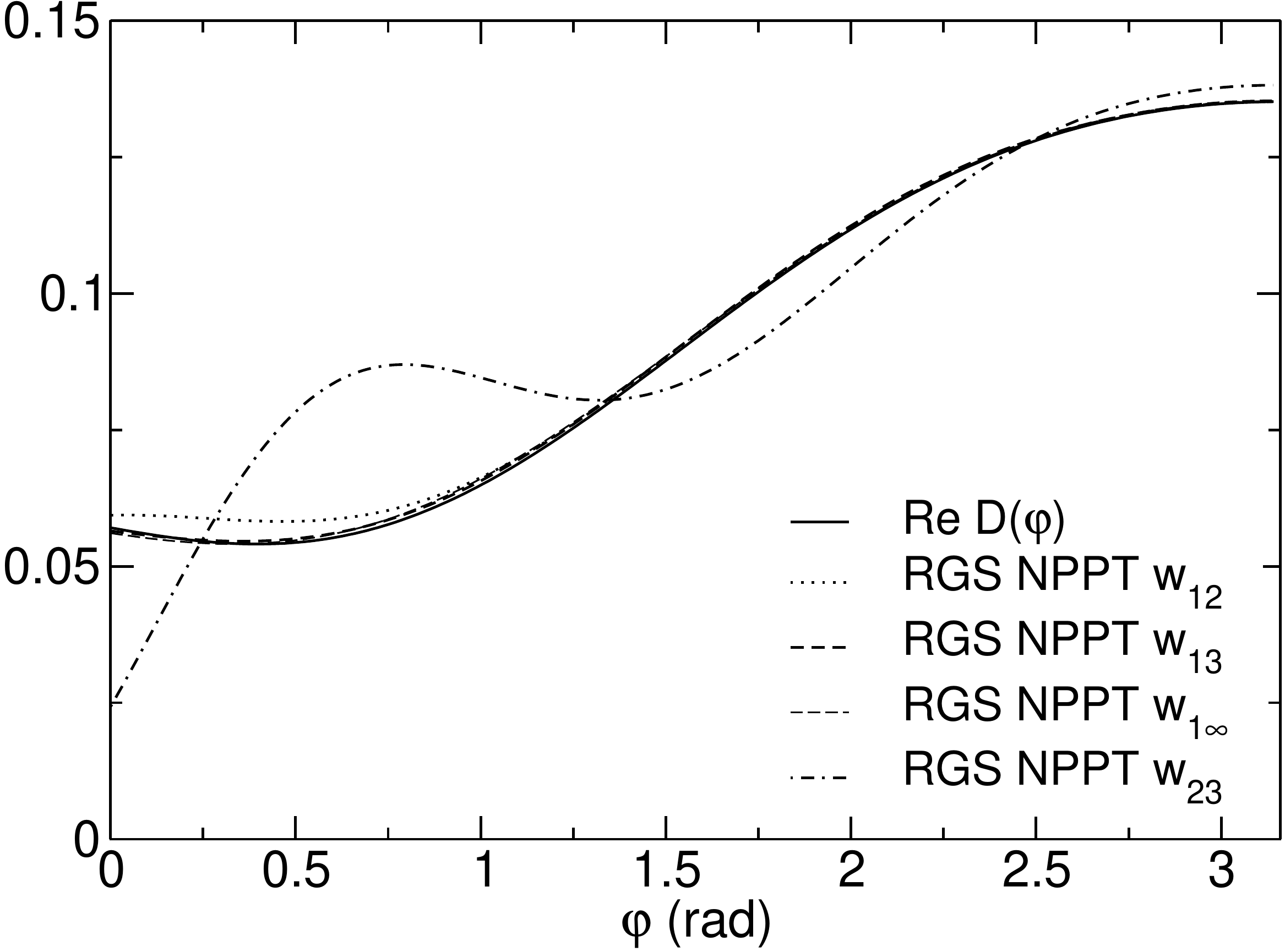}
\caption{Real part of the Adler function for the model (\ref{eq:BBJ}) along the circle $s=m_\tau^2 e^{i\varphi}$, approximated  by   CI and RGS NPPT perturbative expansions defined in (\ref{eq:Dhat}), calculated with N=18 terms, for various conformal mappings $w_{jk}$.  \label{fig:DRCIRG}}\end{center}
\end{figure}

We have so far considered only the CI and the FO expansions. As noted in Refs. \cite{Abbas:2012py,Abbas:2012fi}, the predictions of the RGS expansions are very close to those of the CI expansions, for both standard and improved cases, up to relatively large orders. This is illustrated in Fig. \ref{fig:DRCIRG}, where we show the  real part of the Adler function for the model (\ref{eq:BBJ}) along the circle $s=m_\tau^2 e^{i\varphi}$, and its approximation by CI and RGS non-power expansions calculated with N=18 perturbative terms.  For completeness we present the expansions with all the conformal mappings adopted in section \ref{sec:soft}. The approximations provided by CI and RGS are similar and very good, except for the mapping $w_{23}$. In the latter case, the residual mild 
cut inside the conformal plane $w_{23}$ corresponding to the segment between $u=-1$ and $u=-2$ (see last panel of Fig. \ref{fig:wjk})  limits the convergence radius of the expansion (\ref{eq:prod}). The effect is small at low perturbative orders, but becomes visible at high orders, as is N=18.
 
We shall consider also an integral along the circle $|s|=m_\tau^2$, defined as \cite{BeJa}
\begin{equation}\label{del0}
\delta^{(0)}=\frac{1}{2 \pi i}\!\! \oint\limits_{|s|=m_\tau^2}\!\! \frac{d s}{s}
\left(1- \frac{s}{m_\tau^2}\right)^3 \left(1+\frac{s}{m_\tau^2}\right) \wh D(s).
\end{equation}
As we shall show  in the next section, this quantity enters the theoretical calculation of the $\tau$ hadronic decay width. For the model (\ref{eq:BBJ}),  the exact value of $\delta^{(0)}$, obtained with the Adler function $\wh D(s)$ calculated using Eq. (\ref{eq:Dhat}) with $\alpha_s(m_\tau^2)=0.34$, is   $\delta^{(0)}_{\rm exact}=0.2371 \pm 0.0060$,
where the error is an estimate of the  prescription ambiguity  (cf. Eq. (6.3) of  \cite{BeJa}).

\begin{figure}[htb]\vspace{0.5cm}
\begin{center} \includegraphics[width =6cm]{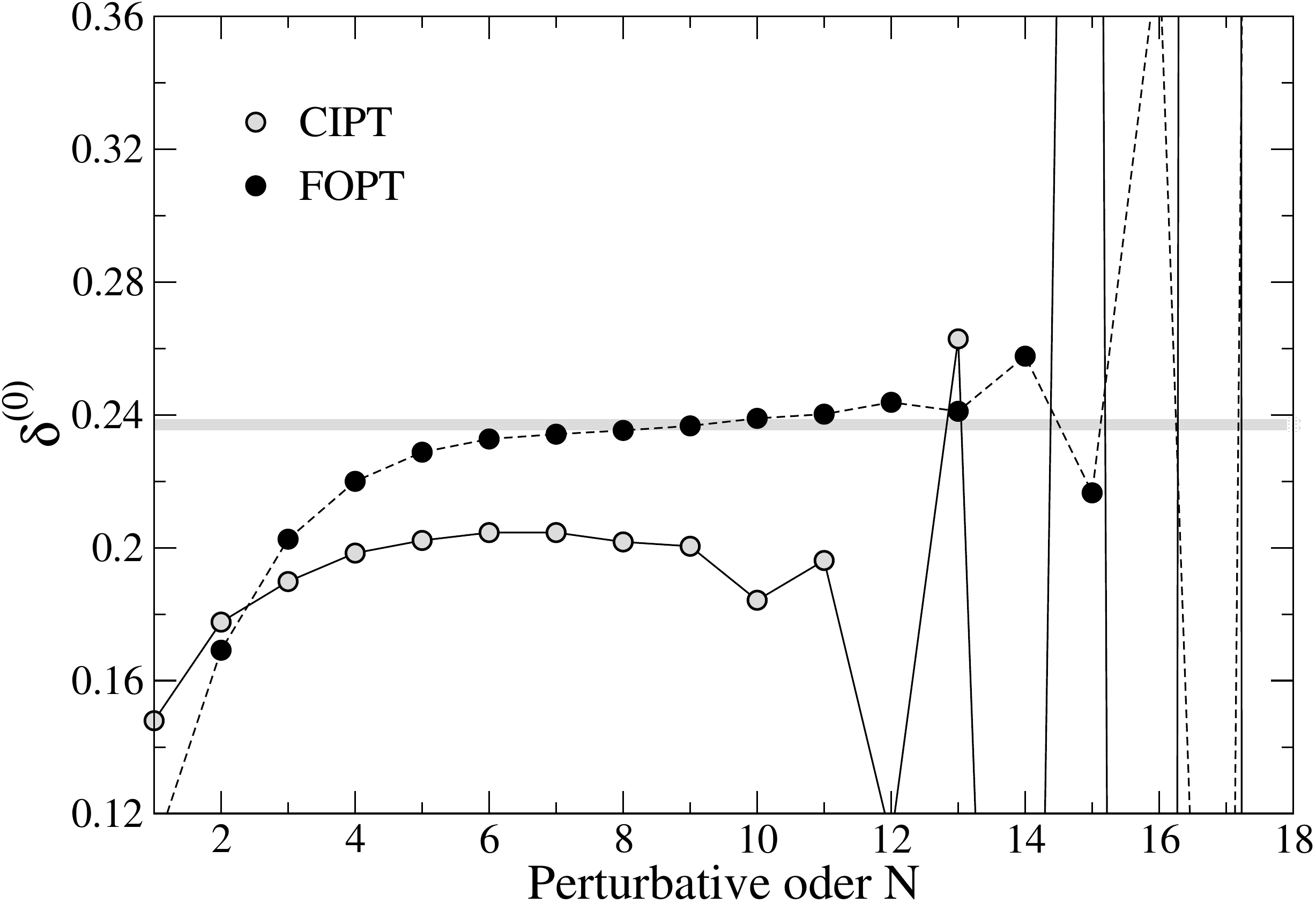}\hspace{1.5cm}\includegraphics[width =6cm]{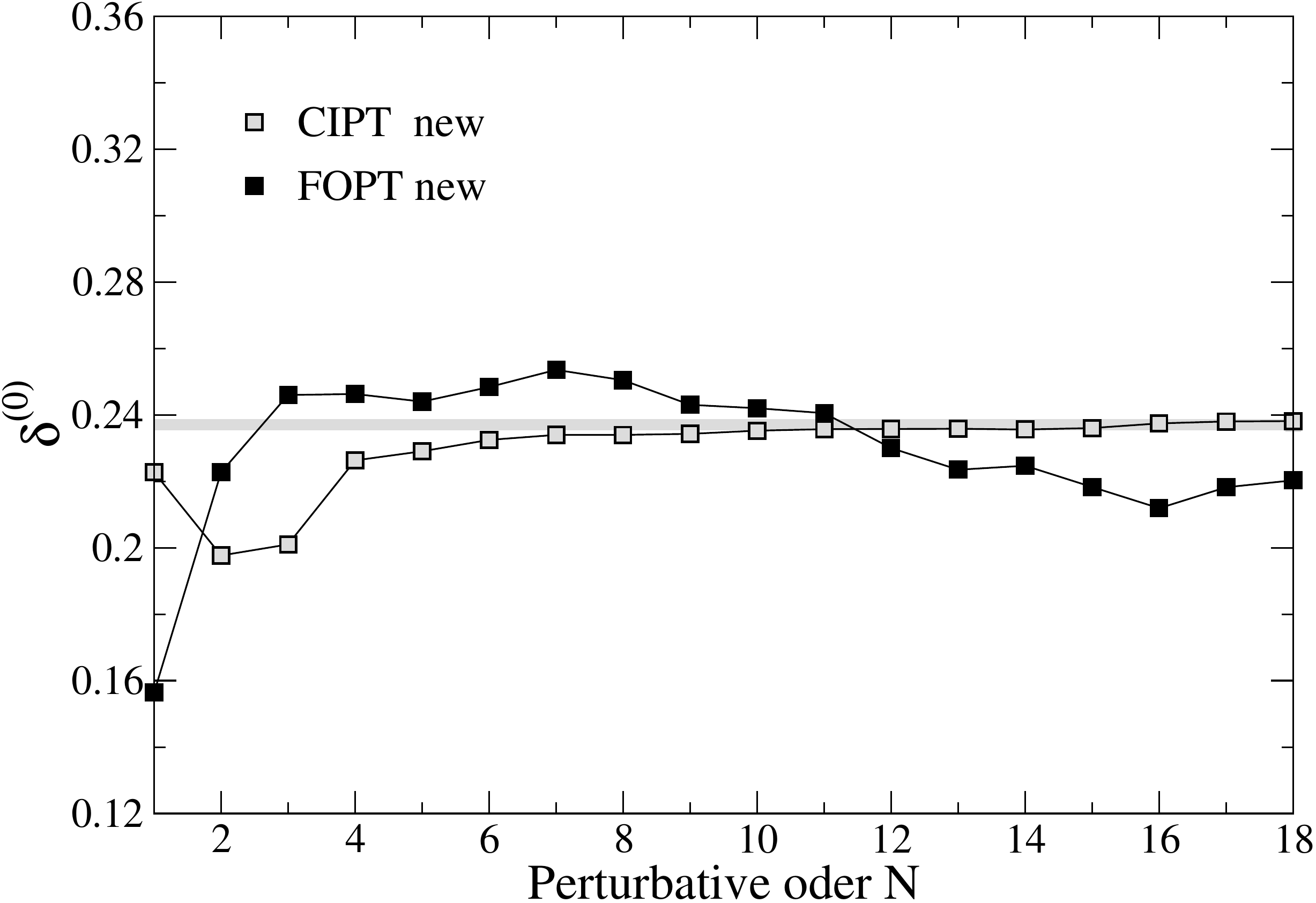}
\caption{Left panel: $\delta^{(0)}$ calculated with the standard CIPT and FOPT expansions as a function of perturbative order for the model (\ref{eq:BBJ}).  The horizontal band is the exact value. Right panel: $\delta^{(0)}$ calculated  with the new  expansions defined in (\ref{eq:Dhat}), for the optimal conformal mapping. \label{fig:delta0}}\end{center}
\end{figure}

In Fig. \ref{fig:delta0}, following Ref. \cite{CaFi2009},  we compare the exact value with the perturbative calculations in the standard CIPT and FOPT, as well that the ``new'' non-power expansions  (\ref{eq:Dhat}) with the optimal conformal mapping (\ref{eq:w}). From the right panel one can see the very good convergence of the CI expansions improved by the optimal conformal mapping of the Borel plane.

More generally,  the ``moments'' $\delta^{(0)}_{i}$ of the spectral function, which can be written as the integrals 
\begin{equation}\label{eq:deli}
\delta^{(0)}_{i}(s_0)= \frac{1}{2\pi i} \!\!\oint\limits_{|s|=s_0}\!\! \frac{d s}{s} \rho_i(s) \wh D(s),
\end{equation}
 have been studied in several papers \cite{BBJ, Boito2013, Abbas:2013usa, Caprini:2013bba}. Here $\rho_i(s)$ is a suitable weight which generalizes the kinematical weight in (\ref{del0}), and  the parameter $s_0$ was set to $m_\tau^2$ or to  lower values. Detailed studies of a large class of moments have been performed in the literature, for the model (\ref{eq:BBJ}) and for other toy models.

Several conclusions follow from the numerical tests performed for the model (\ref{eq:BBJ}). First, we recall that the non-power FO  expansions give a very good description near the euclidian axis, while near the timelike axis the series have poor convergence due to the large imaginary part of the factors $\ln(-s/m_\tau^2)$ present in the coefficients. This implies that renormalization-group summation and a tamed large-order  behaviour are both necessary for a good description of the QCD correlators in the complex $s$ plane. Indeed, the non-power CI and RGS expansion give  very good  approximations up to  high orders for the Adler function and a large class of moments  of the spectral function.

 Similar results are obtained for other toy models, although the detailed behaviour at low orders can be slightly different. From these studies, we conclude that perturbation theory improved by renormalization-group summation in the CI or RGS versions  and the series acceleration by conformal mappings of the Borel plane provides the best description of the physical QCD correlators.

\section{$\alpha_s$ from Hadronic Decays of the $\tau$ Lepton}
\label{sec:tau}
 The hadronic decay width of the $\tau$ lepton has been proposed since a long time \cite{Narison:1988ni, Braaten:1991qm, Pivovarov:1991rh,dLP1,dLP2} as a clean way for
the determination of the strong coupling constant  $\alpha_s$ at a relatively low scale, equal to the mass $m_\tau=1.78 \gev$.
It is convenient to define the ratio $R_\tau$ as:
\be
R_\tau = \dfrac{\Gamma (\tau^- \rightarrow \text{hadrons} \ \nu_\tau)}{\Gamma(\tau^- \rightarrow e \bar\nu_e \nu_\tau)},
\label{eq1}
\ee
where the total decay width $\Gamma (\tau^- \rightarrow \text{hadrons}  \ \nu_\tau)$  is obtained by integration over the invariant mass squared of the final hadron spectrum
\bea
 \Gamma (\tau^- \rightarrow \text{hadrons}\ \nu_\tau)=\int_0^{m_{\tau}^2}  \dfrac{d\Gamma(\tau^- \to \text{hadrons} \ \nu_\tau)}{ds} \ ds.
\label{eq2}
\eea
In the absence of strong and electroweak radiative corrections, the naive prediction for $R_\tau$ is the parton-model value  determined by the color factor $N_c =3$.

\begin{figure}[htb]\vspace{0.5cm}
\begin{center} \includegraphics[width =8cm]{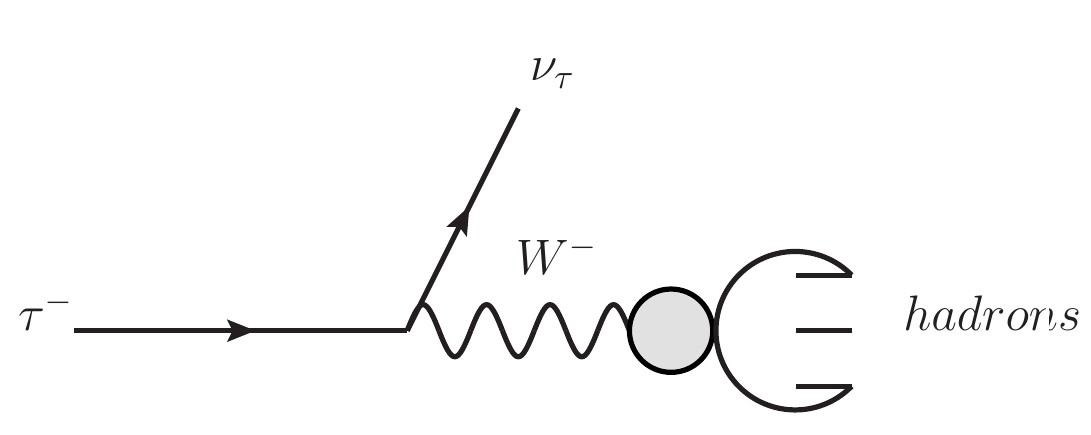}
\caption{The decay  $\tau^-\to {\rm hadrons}\ \nu_\tau$.\label{fig:tau}}\end{center}
\end{figure}

The differential decay width $d\Gamma (\tau^- \rightarrow \text{hadrons}  \ \nu_\tau)/ds$ is calculated from the diagram in Fig. \ref{fig:tau}.   After performing integration over the phase space and using unitarity, one obtains
\be\label{eq:Rtau}
 \Gamma (\tau^- \rightarrow {\rm hadrons}\ \nu_\tau) \sim 12 \pi \int_0^{m_\tau^2} \frac{ds}{m_\tau^2} \left(1- \dfrac{s}{m_\tau^2}   \right)^2 \left( 1+ \frac{2 s}{m_\tau^2}  \right) \text{Im}\Pi (s),
\ee
where $ \Pi(s)$ is the polarization function defined in (\ref{Pi}) and illustrated in Fig. \ref{fig:Pi}.

A straightforward evaluation of (\ref{eq:Rtau}) by perturbative QCD is not possible, because the integral involves a kinematical region at small $s$ on the timelike axis, where the description in terms of free quarks and gluons is not valid. The problem can be handled however by using the analyticity properties of the function $\Pi(s)$ in the complex $s$ plane. Namely, from the general principles of causality and unitarity valid for the QCD confined theory, it is known that  $\Pi(s)$ is a holomorphic function in the complex $s$-plane, except for a branch cut which extends along the real positive axis for $s\ge 4 m_\pi^2$. The branch point is imposed by unitarity, since a pair of $\pi$ mesons is the state of lowest mass that can be produced by the weak current. In addition,  $\Pi(s)$ is a function of real type, {\em i.e.} it satisfies the Schwarz reflection principle  $\Pi(s^*)=  \Pi^*(s)$. From this property it follows, in particular, that the discontinuity across the cut is related to the imaginary part by
\be
 \text{Im}\,  \Pi(s) = -\dfrac{i}{2} \left[ \Pi(s+ i \epsilon) -   \Pi (s - i \epsilon)      \right].
 \label{impi}
\ee

\begin{figure}[htb]
\begin{center}
 \includegraphics[width = 7cm]{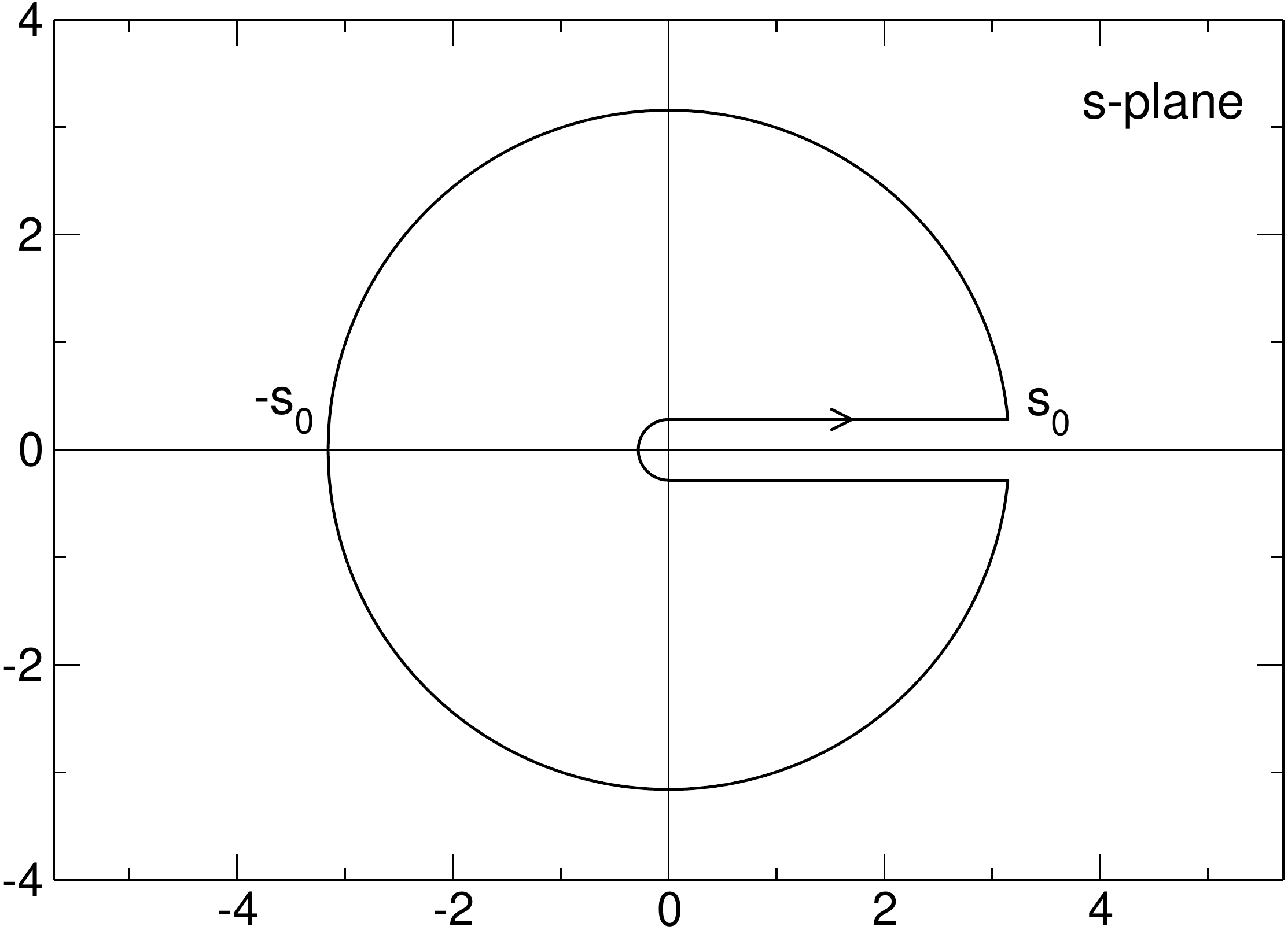}
\caption{Integration contour in the complex $s$ plane where Cauchy theorem is applied to convert an integral along the timelike axis to an integral along the circle. \label{fig:splane}}\end{center}
\end{figure}

Using the analyticity of the exact polarization function  $\Pi(s)$ and applying the Cauchy relation to the   contour of integration shown in Fig. \ref{fig:splane} for $s_0=m_\tau^2$, the integral (\ref{eq:Rtau})  can be written as 
\be\label{eq:Rtaucirc}
 \Gamma (\tau^- \rightarrow \text{hadrons}\ \nu_\tau) \sim 6 \pi i \oint_{|s| = m_\tau^2} \dfrac{ds}{m_\tau^2} \left(1- \dfrac{s}{m_\tau^2}   \right)^2  \left( 1+ \dfrac{2 s}{m_\tau^2}  \right)  \Pi (s).
\ee
Here we have used  the relation (\ref{impi}) and the fact that the remaining factors in the integrand are analytic functions with no discontinuity inside the circle $|s|=m_\tau^2$. After an integration by parts, the integral can be expressed 
in terms of the Adler function $D(s)$, which in turn is written according to (\ref{eq:Dhat}) in terms of the nontrivial QCD contribution $\wh D(s)$. Along the circle $|s|=m_\tau^2$, the function $\wh D(s)$ can be calculated by  perturbative QCD,  which is supposed to be valid for large $|s|$  in the complex plane outside the timelike axis.  Therefore, the r.h.s. of (\ref{eq:Rtaucirc})  is finally related to the quantity $\delta^{(0)}$ defined in Eq. (\ref{del0}).

By including all the couplings, the ratio $R_{\tau}$ produced by the $V+A$ current contribution can be  written as  \cite{BeJa}
\begin{equation}
\label{RtauVA}
R_{\tau} \,=\, N_c\,S_{\rm EW}\,|V_{ud}|^2\,\biggl[\,
1 + \delta^{(0)} + \delta_{\rm EW}' + \sum\limits_{d\geq 2} 
\delta_{ud}^{(d)} \,\biggr] ,
\end{equation}
where $N_c =3$ is the number of quark colors,
$S_{\rm EW}$  and
$\delta_{\rm EW}'$  are electroweak corrections, $\delta^{(0)}$ is the dominant perturbative QCD correction and  $\delta_{ud}^{(d)}$ denote quark-mass corrections
and contributions of higher-dimensional operators, present in the PC series $\Pi_{\rm PC}$  defined in (\ref{eq:PCseries}). 

As shown in \cite{BeJa}, the (less-known) higher terms in the OPE bring a very small  contribution to (\ref{RtauVA}). Therefore, from the measured decay width $R_{\tau}$ it is possible to obtain a fairly accurate phenomenological value of the QCD correction $\delta^{(0)}$, which allows further a precise extraction of the strong coupling $\alpha_s(m_\tau^2)$. The problem has been investigated in many recent papers (see for instance \cite{ BeJa, Davier2008, Pich_Muenchen, CaFi2009, CaFi2011, Abbas:2012fi, BBJ, Boito:2014sta}  and references therein).  It turns out that the  ambiguities related to the renormalization-group summation and the truncation of the perturbative series represent the major part of the  theoretical uncertainty. In view of the discussion in this  presentation,  the expansions improved   by renormalization-group invariance and analytic continuation of the Borel plane should provide the best value of the strong coupling.  Here we quote only the result obtained in \cite{CaFi2009, CaFi2011, CaFi_Manchester, Abbas:2012fi}:
\be\label{eq:alphas}\alpha_s(m_\tau^2)= 0.3192~^{+ 0.0167}_{-0.0126},\ee and refer to the original works  for details of the derivation.

Using the renormalization-group evolution determined from (\ref{eq:rge}), one can translate the value (\ref{eq:alphas}) to the standard scale equal to the mass of $Z$ boson ($m_Z=91.2 \gev$). This gives
$\alpha_s(m_Z^2
)= 0.1184~^{+0.0019}_{-0.0011}$, close to the most recent world average, $\alpha_s(m_Z^2
)= 0.1181 \pm 0.0011$, quoted by Particle Data Group  \cite{PDG}.

\section{Conclusion and Discussion}
\label{sec:summary}

Perturbative QCD is a very successful theory which explains a large number of hadronic observables measured in high-energy processes. Most remarkable is  the consistency between the values of the strong coupling $\alpha_s$ extracted from different observables covering a wide range of energy scales. In recent years, impressive progress  has been achieved by many calculations performed to NNLO and NNLL, or even to higher orders (up to five loops) for some particular quantities. 

In the same time, perturbation theory in QCD is known to be affected by some nontrivial problems.  Thus, the expansions in powers of $\alpha_s$ are divergent series, the coefficients exhibiting a factorial growth at large orders.  Although the observables must be renormalization-group invariant, their truncated series depend on the renormalization scheme and scale.  The truncated series do not have the analyticity properties imposed by causality and unitarity to  the physical correlators  in momentum plane: instead of branch points at the opening of hadronic channels, they possess branch points due to quarks and gluons, and sometimes also unphysical Landau singularities. Finally, the perturbative expansions can not be applied in a straightforward way on the timelike axis in the energy plane, where measurements of the hadronic processes are done. An analytic continuation in the momentum plane, from euclidean or complex values, where the expansions are meaningful,  to the minkowskian regions where hadrons live, is necessary for comparison with experiment.

These problems are correlated to some extent, for instance the renormalization scheme and scale dependence of the truncated expansions is amplified by the growth of the perturbative coefficients.  These difficulties are expected to have small effects at very high energies, where the coupling is very small due to asymptotic freedom, but become visible in applications of perturbative QCD at moderate energies, of a few GeV, such at the mass $m_\tau$ of the $\tau$ lepton. 

In this contribution we have considered the first two of the issues listed above, with emphasis on the fact that  the QCD perturbative series have a zero radius of convergence in the coupling plane. The main idea that we advocated is to use a conformal mapping of the Borel plane and an expansion of the Borel transform in powers of the corresponding variable, in order to perform the analytic continuation of this function outside the domain of convergence of its standard expansion. According to known mathematical results, this technique also improves the asymptotic rate of convergence of the series  in the original region of convergence. In fact, an optimal conformal mapping can be defined, which achieves the best rate of convergence: it is the transformation which maps the whole analyticity domain of the Borel transform onto a disk of radius equal to unity. 

By applying this technique, we have defined a new perturbative expansion of the QCD correlators, in terms of a set of new expansion functions, which replace the standard powers of $\alpha_s$.  When reexpanded in powers of $\alpha_s$,  the new expansions reproduce the perturbative coefficients   known from Feynman diagrams. The new expansion functions have remarkable properties:  they are singular at $\alpha_s=0$ and their expansions  in powers of $\alpha_s$  are divergent. Moreover, they are defined by Laplace-Borel integrals that require a prescription.  This means that the expansion functions resemble the expanded function, {\em i.e.} the QCD correlator, in several of its fundamental features. Therefore, a tamed divergent pattern of the new, non-power expansion of the QCD correlators  is expected. 

These issues have been discussed in detail in the previous sections. We have in particular confirmed the good convergence properties of the new expansions in the case of the Adler function in the complex energy plane, if renormalization-group summation is simultaneously performed. 

The new expansions that we advocate here have also conceptual implications. We recall that in the standard expansions  the factorial growth of the coefficients and the intrinsic ambiguity produced by the IR renormalons are connected and cannot be disentangled. On the other hand, in the new expansion the growth of the coefficients is much tamed, the series being shown to converge if some conditions are met. Thus, the remaining ambiguity,  produced by the infrared regions of the Feynman diagrams, is separated from the divergence of the series and  appears to be a genuine effect. 

This separation can have nontrivial implications on the additional terms present in the expansion (\ref{eq:OPE}) of QCD correlators in the frame of operator product expansion.  According to the modern views on resurgence and the associated trans-series \cite{MStrans}, the existence of the additional series  (\ref{eq:PCseries}) is related to the ambiguities of the dominant perturbative part $\Pi_{\rm pert}(s)$.

There are strong arguments that the series (\ref{eq:PCseries})  itself is 
actually a divergent series. This implies, according to the same ideas of resurgence, the presence of other, additional terms  in the expansion of QCD correlators, beyond the operator product expansion.  According to standard terminology \cite{Blok, Shif1}, these terms, which are exponentially small at large energies, are said to violate quark-hadron duality.  The study of these problems aims to bring clarifications on the application of perturbative QCD to the description of hadronic phenomena at moderate energies.

The summary of the results obtained by research over the last couple of decades  points
to the fact that solutions in quantum field theory require a variety of approaches  beyond the computation of multi-loop Feynman diagrams.  These must go
hand in hand with an analysis of what one can learn about perturbation theory in general.  Advanced mathematical techniques of complex-variable theory offer the requisite
tools to carry out these tasks as demonstrated in the work summarized here.

\subsection*{Acknowledgments}
IC acknowledges support from the Ministry of Research and Innovation, Contract PN 16420101/2016. BA is partly supported by the MSIL Chair of the Division of Physical and Mathematical
Sciences, Indian Institute of Science.


\bigskip


\label{lastpage-01}

\end{document}